\documentclass[11pt,a4paper]{article}

\usepackage[T1]{fontenc}
\usepackage[utf8]{inputenc}
\usepackage{amsmath,amssymb,bm}
\usepackage{graphicx}
\usepackage{booktabs}
\usepackage{multirow}
\usepackage{xcolor}
\usepackage{caption}
\usepackage{subcaption}
\usepackage{tikz}
\usetikzlibrary{arrows.meta,positioning,fit,backgrounds}
\usepackage{listings}

  \usepackage[margin=2.4cm]{geometry}
  \usepackage[colorlinks=true,allcolors=blue!55!black]{hyperref}
  \usepackage[numbers,sort&compress]{natbib}
{\bibliographystyle{unsrtnat}}
  \makeatletter
  \gdef\@abstracttext{}\gdef\@keywordstext{}
  \renewcommand{\author}[2][]{\gdef\@author{#2}}
  \newcommand{\affiliation}[2][]{\g@addto@macro\@author{\\[3pt]{\small\itshape #2}}}
  \newcommand{\emailAdd}[1]{\g@addto@macro\@author{\\[3pt]{\small\texttt{#1}}}}
  \renewcommand{\abstract}[1]{\gdef\@abstracttext{#1}}
  \newcommand{\keywords}[1]{\gdef\@keywordstext{#1}}
  \newcommand{\arxivnumber}[1]{}
  \newcommand{\note}[1]{}
 
  \newcommand{\acknowledgments}{\section*{Acknowledgments}}

\newcommand{\Fobj}{\mathcal{F}}                 % the scan objective (nLL)

\newcommand{\dmsq}[1]{\Delta m^{2}_{#1}}
\newcommand{\ssq}[1]{\sin^{2}\theta_{#1}}
\newcommand{\code}[1]{\texttt{#1}}
\newcommand{\MZp}{M_{Z'}}
\newcommand{\gBL}{g_{BL}}
\newcommand{\vBL}{v_{BL}}
\newcommand{\dCP}{\delta_{\rm CP}}
\newcommand{\TeV}{\,\mathrm{TeV}}
\newcommand{\GeV}{\,\mathrm{GeV}}
\newcommand{\eV}{\,\mathrm{eV}}

\lstdefinestyle{yaml}{
  basicstyle=\ttfamily\footnotesize,
  breaklines=true,
  breakatwhitespace=false,
  columns=fullflexible,
  keepspaces=true,
  frame=single,
  framesep=3pt,
  framerule=0.4pt,
  rulecolor=\color{black!35},
  backgroundcolor=\color{black!3},
  showstringspaces=false,
  xleftmargin=2pt,
  xrightmargin=2pt,
}
\lstnewenvironment{yaml}{\lstset{style=yaml}}{}
\lstnewenvironment{snippet}{\lstset{style=yaml}}{}

\title{\boldmath BSMScanner: a lightweight framework for
beyond-the-Standard-Model global fits}

\author[a]{Mohamed Belfkir}

\affiliation[a]{University of Doha for Science and Technology (UDST), Doha, Qatar}

\emailAdd{mohamed.belfkir@udst.edu.qa}

\abstract{%
We present BSMScanner, a framework in which a beyond-the-Standard-Model (BSM)
parameter scan is specified as a set of declarative YAML documents rather than
written as a program. The user declares the free parameters and their priors,
the mass matrices and the diagonalisations that act on them, the observables,
the theory-validity conditions and the likelihood terms. The framework
validates this specification, resolves it into a dependency graph and compiles
it once into a typed C++ evaluation kernel, which is then driven by an
independently chosen scanning engine: uniform sampling, two
differential-evolution engines, or a staged basin-discovery orchestrator,
optionally followed by Markov-chain posterior refinement. Flavour physics matrix
diagonalisation, PMNS and CKM construction in fixed conventions, tabulated
oscillation likelihoods and a library of common flavour observables are
provided by the framework and imported by name, so that a model file contains only model-specific definitions; quantities without a closed form, such as
a relic density, enter through a C++ or Python plugin interface. We describe
the specification schema, the observable and likelihood library, the scanning
engines and their configuration, and demonstrate the framework on six
benchmark models with $4$ to $26$ free parameters, giving for each the
specification, the scan configuration, the resulting best fit and analytic
cross-checks of the result.}

\date{}
\begin{document}
\maketitle
\flushbottom

%%%%%%%%%%%%%%%%%%%%%%%%%%%%%%%%%%%%%%%%%%%%%%%%%%%%%%%%%%%%%%%%%%%%%%%%%%%%%%%
\section*{Program summary}
\label{sec:summary}

\begin{description}\itemsep2pt
\item[Program title:] BSMScanner (\code{bsm-scanner} on PyPI), version
 0.1.7.

\item[Programming languages:] Python ($\geq$\,3.9), C++17, optional Fortran.
\item[External dependencies:] Eigen~\cite{eigenweb}, NumPy \cite{Harris_2020}, SciPy~\cite{Virtanen:2019joe};
  optional: \code{emcee}~\cite{Foreman-Mackey:2013}, micrOMEGAs~\cite{Belanger:2018ccd}.
\item[Repository:] \url{https://github.com/mbelfkir/bsmscanner} 
\item[Nature of problem:] Phenomenological studies of BSM models require
  repeated evaluation of model predictions --- frequently involving the
  diagonalisation of mass matrices and the extraction of mixing parameters in a
  fixed convention --- inside a numerical search of a high-dimensional parameter
  space subject to experimental likelihoods and theory-validity conditions. The
  corresponding code is usually rewritten for each model, which makes it
  difficult to reuse, to reproduce, and to change the search strategy.
\item[Solution method:] The model, the likelihood and the scan configuration are
  declared in YAML, validated against a schema, resolved into a dependency graph
  and compiled once into an immutable C++ evaluation plan. Scanning engines
  (random sampling, differential evolution, clustered basin search) drive this
  plan through a common interface. Without requesting any coding skills.
\end{description}

%%%%%%%%%%%%%%%%%%%%%%%%%%%%%%%%%%%%%%%%%%%%%%%%%%%%%%%%%%%%%%%%%%%%%%%%%%%%%%%
\section{Introduction}
\label{sec:intro}
Almost every quantitative statement in BSM phenomenology rests on a parameter
scan. A model supplies a Lagrangian with free parameters, a set of measurements
supplies constraints, and a search procedure explores the parameter space to
establish whether the two can be reconciled, where the best fit lies and which
regions survive. The pattern is common to radiative neutrino-mass models,
extended scalar sectors, effective field theories and simplified dark-matter
models.
\\
The infrastructure that performs the search is, however, usually written anew
for each study. Parameter ranges, mass matrices, diagonalisation routines,
observable definitions, the likelihood and the search loop are fused into a
single program written for one model. This has four consequences:
\begin{enumerate}
  \item \emph{Machinery is rewritten rather than reused.} Constructing the PMNS
    matrix from a Majorana mass matrix and extracting the mixing angles and
    phases in a fixed convention are model-independent operations, yet they are
    implemented independently, and with subtly different conventions, in many
    private codes.
  \item \emph{Published scans are difficult to reproduce,} because priors,
    ranges, validity cuts and search settings survive only as choices embedded
    in code.
  \item \emph{The evaluation rate limits the study.} An interpreted inner loop
    caps the number of points that can be evaluated, and studies are designed
    around that cap rather than around the physics question.
  \item \emph{The search algorithm is rarely questioned.} Because it is fused
    to the model, changing from random sampling to differential evolution means
    rewriting the driver, so one algorithm is typically inherited from a
    previous study irrespective of whether it suits the likelihood at hand.
\end{enumerate}

BSMScanner addresses the first three directly and the fourth as a consequence.
Its organising principle is that a model is a \emph{document} rather than a
program: a declarative YAML specification of parameters and priors, mass
matrices and their diagonalisations, derived quantities, observables,
theory-validity conditions, likelihood terms and the scan configuration. The
specification is validated against a schema, resolved into a dependency graph
and compiled once into a typed C++ kernel. Every scanning engine drives the
same compiled object through one interface, so that replacing a random scan by
a staged differential-evolution search is an edit of a few lines in the
\code{scan:} section.

General-purpose global-fitting frameworks address parts of this problem.
\textsc{Gambit}~\cite{GAMBIT:2017yxo} is the most complete example: a modular
architecture that connects a model to external spectrum generators,
relic-density calculators and collider-recasting tools through a
dependency-resolution engine, with a dedicated scanning module,
ScannerBit~\cite{GAMBIT:2017qxg}. HEPfit~\cite{DeBlas:2019ehy} and
flavio~\cite{Straub:2018kue} combine electroweak and flavour observables with a
fit engine behind a common interface. The strength of these tools is coverage:
a model that needs a full spectrum, a relic density and a collider recast can
obtain all three from validated backends, and BSMScanner does not compete with
that. The cost of coverage is architectural weight. Adding a model to
\textsc{Gambit} requires writing a C++ module and registering it with the
dependency-resolution system --- an appropriate cost when the payoff is access
to many external codes, and a disproportionate one when the model's physics
content is a $3\times3$ texture. In flavour model building and EFT
phenomenology this is the common case: there is no spectrum to generate, and
the expensive operation is diagonalising a small matrix $\mathcal{O}(10^6)$
times.

BSMScanner targets this class of models and is deliberately narrower than
\textsc{Gambit} in exchange for being lighter to adopt. No module is written
and no backend registered: the reusable library supplies the PMNS and CKM
construction~\cite{Pontecorvo:1957qd,Maki:1962mu,Cabibbo:1963yz,Kobayashi:1973fv},
the Takagi and singular-value diagonalisations, the standard mixing-angle and
phase conventions, and tabulated oscillation likelihoods, and four scanning
engines sit behind one interface. External calculations remain possible
through a narrow plugin interface, in compiled C++ or in Python, so that an
observable without a closed form in the model parameters --- for example a
relic density or a direct-detection rate from
micrOMEGAs~\cite{Belanger:2018ccd} --- is bound into the same declarative
specification as a named plugin call, without the framework core knowing
anything about the backend.

The paper is organised as follows. Section~\ref{sec:framework} describes the
architecture, the life cycle of a point evaluation, the output artefacts and
the evaluation cost. Section~\ref{sec:model} is the reference for writing a
model: parameters, matrices and diagonalisations, theory checks, observables,
likelihoods, the statistical conventions of the objective, and plugins.
Section~\ref{sec:engines} describes the scanning engines and their
configuration. Section~\ref{sec:examples} applies the full pipeline to six
benchmark models and collects analytic cross-checks of the results, and
section~\ref{sec:conclusions} concludes. Appendix~\ref{app:availability}
summarises installation and the command-line interface.

%%%%%%%%%%%%%%%%%%%%%%%%%%%%%%%%%%%%%%%%%%%%%%%%%%%%%%%%%%%%%%%%%%%%%%%%%%%%%%%
\section{Framework description}
\label{sec:framework}

\subsection{Architecture}

\begin{figure}[t]
\centering
\begin{tikzpicture}[
  font=\small,
  box/.style={draw, rounded corners=1pt, align=center, minimum height=0.85cm, inner sep=4pt, fill=black!3},
  lib/.style={draw, rounded corners=1pt, align=center, minimum height=0.85cm, inner sep=4pt, fill=blue!4, dashed},
  engine/.style={draw, rounded corners=1pt, align=center, minimum height=0.7cm, inner sep=3pt, fill=orange!8},
  arr/.style={-{Latex[length=2mm]}, thick},
  node distance=6mm and 8mm
]
  \node[box, minimum width=3.6cm] (yaml) {YAML model specification\\{\scriptsize parameters, matrices, observables,}\\{\scriptsize theory checks, likelihoods, scan}};
  \node[lib, above=of yaml, minimum width=3.2cm] (corelib) {Reusable \code{core:}\\physics library};
  \node[box, below=of yaml, minimum width=3.6cm] (frontend) {Python front end\\{\scriptsize schema validation, function expansion,}\\{\scriptsize dependency graph, cycle detection,}\\{\scriptsize active-subgraph selection, lowering}};
  \node[box, below=of frontend, minimum width=3.6cm, fill=black!6] (core) {Compiled C++ model\\{\scriptsize typed point evaluator, diagonalisation,}\\{\scriptsize theory checks, likelihood accumulation}};
  \node[lib, right=14mm of core, minimum width=2.6cm] (plugin) {Python/C++ plugins\\{\scriptsize e.g.\ micrOMEGAs}};

  \node[engine, below=12mm of core, xshift=-3.3cm] (rand) {\code{serial\_random}};
  \node[engine, right=3mm of rand] (descipy) {\code{de\_scipy}};
  \node[engine, right=3mm of descipy] (adiver) {\code{adaptive\_diver}};
  \node[engine, right=3mm of adiver] (basin) {\code{basin\_scan}};

  \node[box, below=12mm of descipy.south east, anchor=north, minimum width=7.2cm, fill=black!6] (post) {posterior refinement (\code{emcee}) \quad$\to$\quad statistics post-processing};
  \node[box, below=of post, minimum width=7.2cm] (out) {\code{points.csv} \ \code{best\_fit.json} \ \code{metadata.json} \ \code{summary.json}};

  \draw[arr] (corelib) -- (yaml);
  \draw[arr] (yaml) -- (frontend);
  \draw[arr] (frontend) -- (core);
  \draw[arr] (plugin) -- (core);
  \draw[arr] (core.south) -- ++(0,-4mm) -| (rand);
  \draw[arr] (core.south) -- ++(0,-4mm) -| (descipy);
  \draw[arr] (core.south) -- ++(0,-4mm) -| (adiver);
  \draw[arr] (core.south) -- ++(0,-4mm) -| (basin);
  \draw[arr] (rand.south) |- ++(0,-6mm) -| (post.north west);
  \draw[arr] (descipy.south) |- ++(0,-6mm) -| (post.north);
  \draw[arr] (adiver.south) |- ++(0,-6mm) -| (post.north);
  \draw[arr] (basin.south) |- ++(0,-6mm) -| (post.north east);
  \draw[arr] (post) -- (out);
\end{tikzpicture}
\caption{The BSMScanner pipeline. A YAML specification, optionally importing
framework-owned \code{core:} blocks, is validated and compiled once into a typed
C++ model; plugins admit external calculations. Any of the four scanning
engines drives the same compiled object, and the optional posterior and
statistics stages act on the completed scan.}
\label{fig:architecture}
\end{figure}
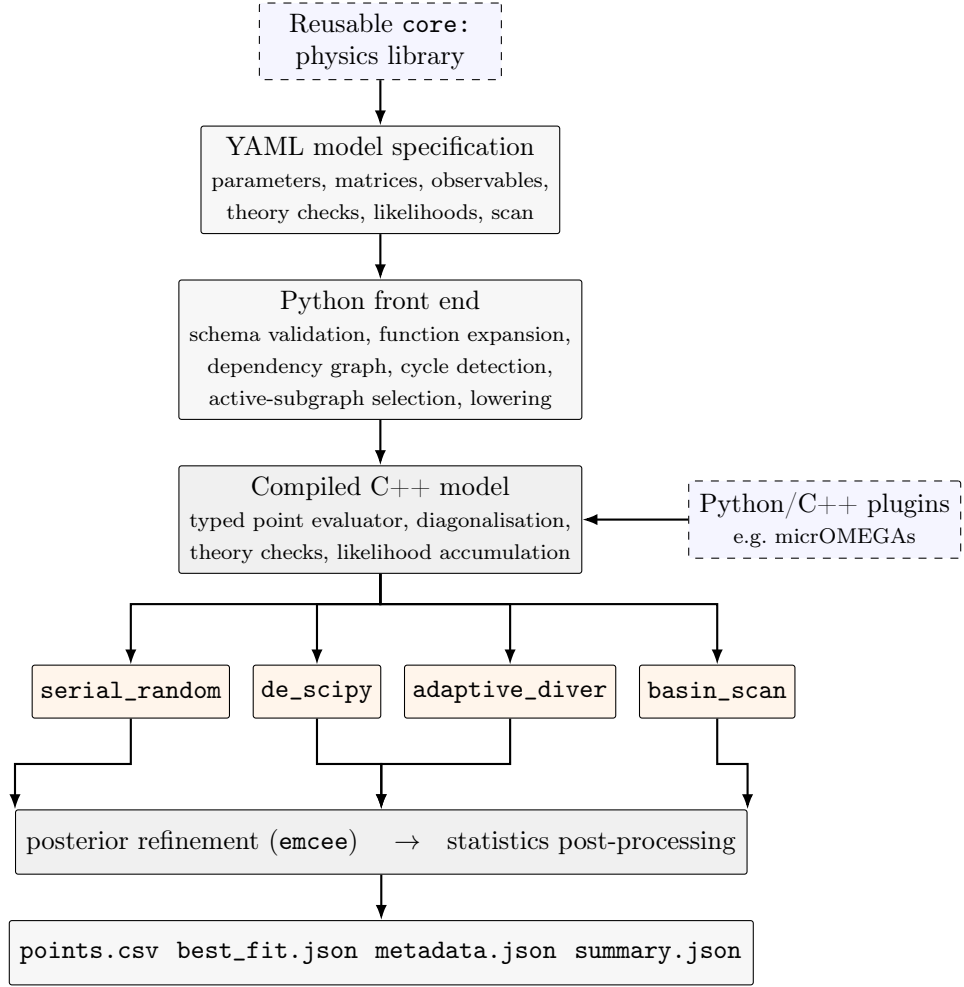

BSMScanner consists of three layers. A Python front end reads the model files,
validates them against the schema, expands reusable analytic functions, builds
a typed dependency graph, detects cycles, selects the subgraph needed by the
requested outputs and likelihood terms, and lowers the result to a compact
execution plan. A C++ core owns the immutable compiled plan and evaluates points
without calling back into Python: it caches node values within a point,
performs matrix diagonalisation with Eigen~\cite{eigenweb}, separates theory
checks from observable and likelihood evaluation, and exposes a native scan
runner. An optional Fortran layer, used only for numerically sensitive loop
functions and external scanner bridges, has no dependence on the rest of the
framework. Parsing, name resolution, schema validation and function expansion
therefore happen once, at load time; the repeated cost of a scan is the
execution of a pre-compiled instruction stream. Figure~\ref{fig:architecture} shows the pipeline. A model is assembled from a
top-level manifest that imports YAML fragments: the user's own files and
framework-owned blocks referenced with a \code{core:} prefix, which resolves
wherever the package is installed. The front end merges the fragments,
validates every section, expands \code{functions:}, builds the dependency graph
and computes the active subgraph. The graph is lowered into a
\code{CompiledModelSpec} and passed to the C++ core, which constructs the
immutable \code{CompiledModel}. From this point a scan engine is a driver that
proposes parameter vectors, calls the compiled evaluator and records the
results; optional posterior-refinement and statistics stages act on the
completed scan without altering the model.

\subsection{Point evaluation and validity}
\label{sec:lifecycle}

The evaluation of a point is the same whichever engine requests it. The
external parameters are populated in declaration order; the active derived
nodes are evaluated in topological order; each theory check is evaluated as
soon as its dependencies are available and can terminate the point early with a
deterministic failure reason; the observables are evaluated; the likelihood
terms are accumulated into a per-term negative log-likelihood; and only the
requested outputs are materialised. Each record carries a technical status
(\code{ok}, \code{missing\_input}, \code{invalid\_point}, \code{numerical\_error},
\code{evaluation\_error}, \code{non\_finite\_objective}) and a physical
\code{valid} flag; a point that fails a theory check is a normal outcome, with
technical status \code{ok} and \code{valid = false}, and the failure reasons are
counted in \code{summary.json}. Downstream statistics must use \code{valid}, not
the technical status. Because parameter ordering is fixed by declaration order
and recorded in \code{metadata.json}, a run can be reconstructed exactly from
its artefacts.

\subsection{Output artefacts}

Every run directory contains the four files of Table~\ref{tab:artefacts}. They
are plain CSV and JSON, so that the pipeline is
\emph{engine $\to$ raw and statistical outputs $\to$ external plotting}; the
framework does not produce plots, and a user's analysis code is never coupled
to how the scan was produced.

\begin{table}[t]
\centering\small
\begin{tabular}{lp{10.6cm}}
\toprule
File & Content \\
\midrule
\code{points.csv} & one row per evaluated point: scanned parameters, requested outputs, per-term likelihood contributions, total objective, \code{status}, \code{valid} and failure reason (invalid points optional via \code{save\_invalid\_points}). \\
\code{metadata.json} & model name and version, framework version, run identifier, UTC timestamp, engine, objective mode, seed, parameter order with bounds and priors, resolved engine settings. \\
\code{summary.json} & number of evaluations, saved and valid points, interruption flag, counters per technical status, histogram of failure reasons, engine stop reason. \\
\code{best\_fit.json} & best valid point: parameters, objective, per-term contributions and requested outputs. \\
\bottomrule
\end{tabular}
\caption{Artefacts written by every run.}
\label{tab:artefacts}
\end{table}

\subsection{Evaluation cost}
\label{sec:performance}

Table~\ref{tab:performance} gives the measured cost of an end-to-end
\code{serial\_random} scan --- point generation, compiled evaluation and CSV
output of every point --- for three of the benchmarks of
section~\ref{sec:examples}. The marginal cost per point is obtained from the
difference between runs of $2\times10^3$ and $5\times10^4$ points, which removes
the one-off cost of loading and compiling the model. For the 26-parameter
Scotogenic model, which requires a complex $3\times3$ diagonalisation, PMNS
extraction, six spline-interpolated likelihood tables and nine further terms per
point, the cost is about $50\,\mu$s, so that $10^6$ points take under a minute
on a single core.

\begin{table}[t]
\centering\small
\begin{tabular}{lcccc}
\toprule
Model & Par. & Wall time, $2\times10^3$ pts & Wall time, $5\times10^4$ pts & Marginal cost / point \\
\midrule
Scotogenic         & 26 & 0.31\,s & 2.56\,s & $47\,\mu$s \\
Two-Higgs-doublet  & 6  & 0.21\,s & 1.19\,s & $20\,\mu$s \\
SMEFT Warsaw       & 9  & 0.19\,s & 0.82\,s & $13\,\mu$s \\
\bottomrule
\end{tabular}
\caption{End-to-end \code{serial\_random} throughput, single process, Apple M4
Pro, \code{bsm-scanner}. Wall times include model
loading, compilation and writing every point to \code{points.csv}.}
\label{tab:performance}
\end{table}

%%%%%%%%%%%%%%%%%%%%%%%%%%%%%%%%%%%%%%%%%%%%%%%%%%%%%%%%%%%%%%%%%%%%%%%%%%%%%%%
\section{Model declaration}
\label{sec:model}

A model is one YAML file or a manifest that imports several.
Table~\ref{tab:schema} lists the top-level sections; all are optional except
\code{parameters:} and the combination of \code{matrices:},
\code{observables:} and \code{likelihoods:} that the physics requires. A first
runnable model is obtained with
\begin{snippet}
pip install bsm-scanner
bsm-scanner new-model mymodel
bsm-scanner run --model mymodel/model.yaml --run-dir mymodel/runs/first
\end{snippet}
which creates a small, complete model wired to the core library and scans it,
so that a model is always developed from a working baseline. While writing a
model, \code{bsm-scanner core list} prints the reusable blocks shipped with the
package and \code{bsm-scanner core show <block>} prints the constants,
functions, observables and theory checks a block contributes before it is
imported (Appendix~\ref{app:availability}).

\begin{table}[t]
\centering\small
\begin{tabular}{ll}
\toprule
Section & Contents \\
\midrule
\code{metadata} & name, version, description, free-text tags \\
\code{imports} & the model's own YAML fragments and \code{core:} library blocks \\
\code{python\_plugins} & Python modules registering plugin functions (section~\ref{sec:plugins}) \\
\code{parameters} & scanned and fixed inputs, with ranges, priors and defaults \\
\code{constants} & fixed numerical values referenced by expressions \\
\code{functions} & reusable analytic functions, expanded at load time \\
\code{derived\_scalars}, \code{derived\_complex} & named expressions over parameters and constants \\
\code{matrices} & mass and coupling matrices, with diagonalisation metadata \\
\code{diagonalizations} & explicit diagonalisation nodes (usually generated automatically) \\
\code{mixing\_matrices} & PMNS/CKM-type products of rotation matrices \\
\code{observables} & named physical quantities, expression-, projection- or plugin-backed \\
\code{theory\_checks} & pass/fail validity conditions, separate from the likelihood \\
\code{likelihoods} & the terms summed into the scan objective \\
\code{outputs} & quantities written to \code{points.csv} \\
\code{scan} & engine choice and settings \\
\code{posterior}, \code{statistics} & optional post-scan stages \\
\bottomrule
\end{tabular}
\caption{Top-level sections of the model schema. Named-list sections
(parameters, matrices, observables, likelihoods, \dots) are concatenated across
imported fragments and mapping sections (metadata, outputs, scan) are merged by
key; duplicate names and conflicting keys are rejected at load time.}
\label{tab:schema}
\end{table}

For a model of any size, the recommended layout separates the manifest from the
sectors,
\begin{snippet}
models/example/
  model.yaml          # metadata + imports only
  parameters.yaml     # parameters, ranges, priors
  constants.yaml      # constants
  functions.yaml      # reusable analytic functions
  derived.yaml        # derived scalar and complex quantities
  matrices.yaml       # matrices and diagonalisations
  observables.yaml    # observables
  likelihoods.yaml    # likelihood terms
  outputs.yaml        # saved outputs (parameters are always saved)
  scan.yaml           # engine configuration
\end{snippet}
so that, for example, replacing a normal-ordering neutrino likelihood by an
inverted-ordering one is a one-line change of the manifest. A complete manifest
for a lepton-sector model in normal ordering reads
\begin{yaml}
metadata:
  name: leptontest_normal
imports:
  - parameters.yaml
  - functions.yaml
  - derived.yaml
  - matrices.yaml
  - core:neutrino/normal.yaml
  - likelihoods.yaml
  - outputs.yaml
  - scan.yaml
\end{yaml}
The model never defines the PMNS matrix, the mixing-angle extraction or the
mass-ordering map; these are imported from the library as shown in Table~\ref{tab:corelib}, and the model files contain only the parameters, the
matrices and the choice of active likelihood terms.

\begin{table}[t]
\centering\small
\begin{tabular}{ll}
\toprule
Block & Provides \\
\midrule
\code{core:constants/physics\_constants.yaml} & charged-lepton mass ratios and shared constants \\
\code{core:neutrino/observables\_common.yaml} & PMNS extraction, angles, $\dCP$, $J$, masses, $m_\beta$, $m_{\beta\beta}$ \\
\code{core:neutrino/observables\_\{normal,inverted\}.yaml} & ordering-aware mass mapping \\
\code{core:neutrino/constants\_\{normal,inverted\}.yaml} & reference splittings for mass-scale normalisation \\
\code{core:quark/ckm\_observables.yaml} & CKM elements, angles, $\delta_{\rm CKM}$, $J_{\rm CKM}$, Wolfenstein parameters \\
\code{core:quark/quark\_mass\_ratios.yaml} & ordered light-to-heavy quark mass ratios \\
\code{core:data/nufit/\{Normal,Inverted\}/*.csv} & tabulated NuFIT-6.0 $\Delta\chi^2$ profiles version \cite{Esteban:2024eli}  \\
\bottomrule
\end{tabular}
\caption{Reusable blocks shipped with the package; \code{bsm-scanner core list}
prints the current set.}
\label{tab:corelib}
\end{table}

\subsection{Parameters}
\label{sec:parameters}

Each parameter carries a value type, a scan flag, a range, a default value and a
prior:
\begin{yaml}
parameters:
  - {name: Mdm, value_type: real, scan: true, lower: 50.0, upper: 1500.0,
     default: 685.03, prior: log}
  - {name: sh,  value_type: real, scan: true, lower: 1.0e-3, upper: 0.997,
     default: 0.108, prior: flat}
  - {name: ypr11, value_type: real, scan: true, lower: -3.0, upper: 3.0,
     default: 5.4e-6, prior: signed_log, min_abs: 1.0e-8}
\end{yaml}
The prior shown in Table~\ref{tab:priors} defines the map from an engine's internal
unit-hypercube coordinate to the physical parameter, and therefore what
``uniform sampling'' means for that parameter. It is a physics choice and is
recorded in \code{metadata.json}. Since the engines operate on real vectors, a
complex parameter is declared as two real parameters (real and imaginary parts,
or modulus and phase) and reconstructed by a \code{derived\_complex} node.

\begin{table}[t]
\centering\small
\begin{tabular}{lp{10cm}}
\toprule
Prior & Meaning \\
\midrule
\code{flat} & uniform in the parameter between \code{lower} and \code{upper}. \\
\code{log} & uniform in $\log_{10}$ of the parameter; requires $\code{lower}>0$. Suited to scales spanning decades (masses, decay constants). \\
\code{signed\_log} & uniform in $\log_{10}|x|$ for $|x|\ge\code{min\_abs}$, with an independently sampled sign. Suited to couplings whose sign is physical and whose magnitude spans decades. \\
\code{fixed} & pins the parameter to its default while keeping it addressable by every expression. \\
\bottomrule
\end{tabular}
\caption{Parameter priors.}
\label{tab:priors}
\end{table}

\subsection{Constants, derived quantities and functions}

Fixed numbers are declared once and referenced by name, and named expressions over parameters and constants are declared as
\code{derived\_scalars} or \code{derived\_complex}:
\begin{yaml}
constants:
  - {name: v_sm, value: 246.22}
  - {name: mh_sm, value: 125.25}
derived_complex:
  - {name: he1, value_type: complex, expression: "he1_re + 1j*he1_im"}
derived_scalars:
  - {name: meta_R_sq, expression: "m_eta_sq + 0.5*(lambda3 + lambda4 + lambda5)*v_sm**2"}
  - {name: meta_R, expression: "sqrt(meta_R_sq)"}
\end{yaml}
A repeated analytic form is written once as a \code{function} and inlined at
load time into every expression that calls it, so that the abstraction has no
run-time cost. For example, the one-loop function of the Scotogenic
model~\cite{Ma:2006km},
\begin{equation}
\Lambda_k=\frac{M_k}{16\pi^2}\left[\frac{m_R^2}{m_R^2-M_k^2}\ln\frac{m_R^2}{M_k^2}
-\frac{m_I^2}{m_I^2-M_k^2}\ln\frac{m_I^2}{M_k^2}\right],
\end{equation}
is declared as
\begin{yaml}
functions:
  - name: ma_loop
    args: [MN, mR, mI]
    expression: >
      MN/(16*pi*pi) * (
        (mR**2/(mR**2 - MN**2))*log((mR**2)/(MN**2))
        - (mI**2/(mI**2 - MN**2))*log((mI**2)/(MN**2))
      )
derived_scalars:
  - {name: loop_N1, expression: "ma_loop(MN1, meta_R, meta_I)"}
\end{yaml}
i.e.\ as an ordinary algebraic expression in the model's own file, with no
compiled code involved.

\subsection{Matrices and diagonalisation}
\label{sec:matrices}

Matrices are written entry by entry as expressions and tagged with a semantic
\code{type} from which the diagonalisation method is inferred:
\begin{yaml}
matrices:
  - name: neutrino_mass_matrix
    value_type: complex_matrix
    type: majorana_mass          # -> Takagi factorisation
    role: neutrino
    diagonalize: true
    rows:
      - ["2*Ye*ye*norm", "(Ye*ymu + Ymu*ye)*norm", "(Ye*ytau + Ytau*ye)*norm"]
      - ["(Ye*ymu + Ymu*ye)*norm", "2*Ymu*ymu*norm", "(Ymu*ytau + Ytau*ymu)*norm"]
      - ["(Ye*ytau + Ytau*ye)*norm", "(Ymu*ytau + Ytau*ymu)*norm", "2*Ytau*ytau*norm"]
\end{yaml}
Table~\ref{tab:diagmethods} lists the supported types. A complex symmetric
(Majorana) matrix is brought to diagonal form by the Takagi
factorisation~\cite{Horn:2012matrix},
\begin{equation}
U^{T} M U = \mathrm{diag}(m_{1},m_{2},m_{3}), \qquad m_{i}\ge 0 ,
\label{eq:takagi}
\end{equation}
which fixes the phases of the columns of $U$ and therefore the Majorana phases.
A general complex (Dirac) matrix is diagonalised by the singular-value
decomposition (SVD) $U_{L}^{\dagger} M U_{R} = \mathrm{diag}(m_i)$, and a Hermitian
matrix by a unitary transformation. Note that applying an SVD to a Majorana
matrix yields the correct masses and moduli $|U_{\alpha i}|$, and hence the
mixing angles and $\dCP$, but leaves each column of $U_L$ defined only up to a
phase: quantities that depend on the Majorana phases, such as $m_{\beta\beta}$,
are then not defined, and the Takagi factorisation must be used. If
\code{diagonalize: true} is given without a name, a node
\code{diag\_\_<role>} is generated; output names can also be set explicitly:
\begin{yaml}
matrices:
  - name: Ml
    value_type: complex_matrix
    type: complex_general
    role: charged_lepton_mass
    diagonalize:
      method: svd
      output: {masses: charged_lepton_masses, left_unitary: U_l_L, right_unitary: U_l_R}
    rows: [["0.000511","0.0","0.0"], ["0.0","0.10566","0.0"], ["0.0","0.0","1.77686"]]
\end{yaml}
Since the core operates on roles rather than on fixed names, neutrino-only, quark-only, lepton-sector and full
four-matrix models are ordinary uses of the same mechanism.

\begin{table}[t]
\centering\small
\begin{tabular}{lll}
\toprule
\code{type} & Method inferred & Convention \\
\midrule
\code{complex\_symmetric} / \code{majorana\_mass} & \code{takagi} & $U^{T}MU=D$, $D\ge0$ \\
\code{complex\_general} / \code{dirac\_mass} & \code{svd} / \code{svd\_complex} & $U_L^{\dagger}MU_R=D$, $D\ge0$ \\
\code{hermitian} & \code{hermitian\_eigh} & $U^{\dagger}HU=D$ \\
\code{real\_symmetric} & \code{self\_adjoint\_eigen} & $U^{T}MU=D$ \\
\bottomrule
\end{tabular}
\caption{Diagonalisation method inferred from the declared \code{type}. The
method can also be set explicitly, or through \code{diagonalizations:} entries.}
\label{tab:diagmethods}
\end{table}

Mixing matrices are built from already-diagonalised rotations,
\begin{yaml}
mixing_matrices:
  - {name: PMNS, type: left_mismatch, convention: U_left_dagger_U_right,
     left: U_l_L, right: U_nu, output: U_PMNS}
  - {name: CKM,  type: left_mismatch, convention: U_left_dagger_U_right,
     left: U_u_L, right: U_d_L, output: V_CKM}
\end{yaml}
implementing $U_{\rm PMNS}=U_{l,L}^{\dagger}U_{\nu}$ and
$V_{\rm CKM}=U_{u,L}^{\dagger}U_{d,L}$. A neutrino-only model that requests a
PMNS matrix without declaring a charged-lepton rotation receives
$U_{l,L}=\mathbb{1}$ unless this fallback is disabled; a CKM request always
requires both rotations.

\subsection{Theory checks}

A theory check is a pass/fail condition, evaluated as soon as its dependencies
are available and kept separate from the likelihood:
\begin{yaml}
theory_checks:
  - {name: positive_mphi_sq, condition: "mphi_sq > 0", fatal: true,
     message: "mphi_sq must remain positive"}
  - {name: perturbative_gauge_yukawa,
     condition: "gBL < sqrt(4*pi) and yN1 < sqrt(4*pi)", fatal: true,
     message: "a B-L coupling is non-perturbative"}
\end{yaml}
A failed fatal check invalidates the point before any likelihood term is
evaluated, and its message is counted in \code{summary.json}. Perturbativity,
vacuum stability and EFT validity are statements about whether a point defines
a consistent theory, not about how well it fits data; keeping them out of the
likelihood keeps the objective free of arbitrary penalty scales and makes the
invalid fraction of the prior volume a directly reported quantity.

\subsection{Observables}
\label{sec:observables}

An observable is a named, savable quantity built on parameters, derived nodes
and diagonalisation or mixing-matrix outputs. Three constructions are
available: an expression,
\begin{yaml}
observables:
  - {name: dm21, expression: "m2**2 - m1**2"}
\end{yaml}
a projection of a diagonalisation output,
\begin{yaml}
observables:
  - {name: m1, projection: {from: diag__neutrino, quantity: singular_values, index: 0}}
\end{yaml}
and a plugin call for a quantity computed by an external code
(section~\ref{sec:plugins}):
\begin{yaml}
observables:
  - name: Omega
    plugin_call:
      plugin: relic_backend
      function: omega
      bindings: {Mchi: Mpsi, Mh: mh_ref}
\end{yaml}
Once declared, an observable is available to \code{outputs:},
\code{likelihoods:} and any other expression by name, with no change elsewhere
in the model.

Most flavour observables are not model specific, and the library
as in Table~\ref{tab:corelib} provides them under fixed names and conventions. The
leptonic set follows the PDG parametrisation~\cite{ParticleDataGroup:2024cfk},
\begin{equation}
\ssq{12}=\frac{|U_{e2}|^{2}}{1-|U_{e3}|^{2}}, \qquad
\ssq{13}=|U_{e3}|^{2}, \qquad
\ssq{23}=\frac{|U_{\mu3}|^{2}}{1-|U_{e3}|^{2}} ,
\end{equation}
with the Dirac phase obtained as $\dCP=\operatorname{atan2}(\sin\dCP,\cos\dCP)$,
where $\sin\dCP=J/J_{\max}$ follows from the Jarlskog invariant
$J=\operatorname{Im}(U_{e1}U_{\mu2}U_{e2}^{*}U_{\mu1}^{*})$ and $\cos\dCP$ from
$|U_{\mu1}|^2$, so that the quadrant is resolved. The phase is exported both in
$[0^\circ,360^\circ)$ (\code{deltaCP\_deg\_0\_360}, the NuFIT publication
convention) and in $[-180^\circ,180^\circ]$ (\code{deltaCP\_deg\_m180\_180},
the domain of the shipped $\dCP$ data table). The effective masses are
\begin{equation}
m_{\beta}=\Big(\textstyle\sum_i |U_{ei}|^{2} m_{i}^{2}\Big)^{1/2}, \qquad
m_{\beta\beta}=\Big|\textstyle\sum_i U_{ei}^{2} m_{i}\Big| ,
\end{equation}
the latter requiring the Takagi convention of eq.~\eqref{eq:takagi}. Since many
flavour textures predict only mass ratios, the neutrino blocks also provide the
normalisation of the overall mass scale to a measured splitting as a shared,
explicit operation. The quark-sector block exposes the CKM elements, angles,
$\delta_{\rm CKM}$, $J_{\rm CKM}$ and the Wolfenstein parameters from a node named
\code{V\_CKM}, independently of how that matrix was built.

\subsection{Likelihoods}
\label{sec:likelihoods}

Each likelihood term names one observable (a list for
\code{multivariate\_gaussian}) and contributes a function $f_t$ to the scan
objective,
\begin{equation}
\Fobj(\bm{\theta}) \;=\; \sum_{t} f_{t}\big(O_{t}(\bm{\theta})\big) ,
\label{eq:objective}
\end{equation}
evaluated at points that pass every theory check and reported as \code{nLL}.
The nine kinds are listed in Table~\ref{tab:likelihoods}.

\begin{table}[t]
\centering\small
\begin{tabular}{lll}
\toprule
Kind & Encodes & Contribution $f_t$ \\
\midrule
\code{gaussian}              & $\mu\pm\sigma$                      & $\tfrac12 z^{2}$, \ $z=(O-\mu)/\sigma$ \\
\code{asymmetric\_gaussian}  & $\mu^{+\sigma_+}_{-\sigma_-}$        & $\tfrac12 z^{2}$, $\sigma=\sigma_\pm$ for $O\gtrless\mu$ \\
\code{upper\_limit}          & $O\le b$                             & $0$ if $O\le b$, else $\tfrac12[(O-b)/\sigma]^{2}$ \\
\code{lower\_limit}          & $O\ge b$                             & $0$ if $O\ge b$, else $\tfrac12[(b-O)/\sigma]^{2}$ \\
\code{interval}              & $a\le O\le b$                        & $0$ inside, $\tfrac12(d/\sigma)^{2}$ at distance $d$ outside \\
\code{hard\_cut}             & $a\le O\le b$                        & $0$ inside, point invalid outside \\
\code{table\_lookup}         & tabulated profile $g(O)$             & interpolated $g(O)$, as published \\
\code{multivariate\_gaussian}& correlated measurements              & $\tfrac12\,\delta^{T}C^{-1}\delta$, \ $\delta=O-\mu$ \\
\code{custom}                & anything else                        & plugin callback \\
\bottomrule
\end{tabular}
\caption{Likelihood kinds available under \code{likelihoods:}.}
\label{tab:likelihoods}
\end{table}

Adding a constraint never requires a change to the framework: it is one entry
in \code{likelihoods:} naming an existing observable. A Gaussian measurement
reads
\begin{yaml}
likelihoods:
  - {name: higgs_mass, kind: gaussian, observable: HiggsMass, mean: 125.10, sigma: 0.14}
\end{yaml}
a one-sided limit is a \code{hard\_cut} or an \code{upper\_limit}, depending on
whether the bound is a veto or a soft penalty,
\begin{yaml}
likelihoods:
  - {name: mu_to_e_gamma_limit, kind: hard_cut, observable: br_mu_e_gamma,
     lower: 0.0, upper: 1.5e-13}
\end{yaml}
and a published one-dimensional profile, such as a NuFIT
projection~\cite{Esteban:2024eli}, is imported together with its interpolation
and out-of-range treatment:
\begin{yaml}
likelihoods:
  - {name: dm21_nufit, kind: table_lookup, observable: log10_dm21,
     table_file: core:data/nufit/Normal/dm21.csv,
     interpolation: cubic_spline,
     out_of_range_penalty_scale: 4.0e4, out_of_range_penalty_cap: 1.0e6}
\end{yaml}
Outside the tabulated domain $[x_{\min},x_{\max}]$ the term returns
\begin{equation}
f(O) = g(x_{\rm edge}) + \min\!\big[k\,(O-x_{\rm edge})^2,\;f_{\rm cap}\big],
\end{equation}
with $x_{\rm edge}$ the nearest table edge, $k$ the penalty scale and
$f_{\rm cap}$ the cap. Anchoring the penalty to the boundary value keeps $f$
continuous at the edge.

Correlated measurements are entered as a single \code{multivariate\_gaussian}
term rather than as independent Gaussians, which would neglect the correlation;
for the PDG oblique-parameter fit~\cite{ParticleDataGroup:2024cfk}
($S=-0.04\pm0.10$, $T=0.01\pm0.12$, $\rho=0.93$),
\begin{yaml}
likelihoods:
  - name: oblique_ST
    kind: multivariate_gaussian
    observables: [ObliqueS, ObliqueT]
    means: [-0.04, 0.01]
    covariance: [[0.0100, 0.01116], [0.01116, 0.0144]]
\end{yaml}
A constraint that has no closed form in the model's parameters and
diagonalisation outputs --- a direct-detection $p$-value from an external code,
for example --- is a \code{custom} term routed through the same
\code{plugin\_call} mechanism as plugin-backed observables:
\begin{yaml}
likelihoods:
  - name: direct_detection_pvalue
    kind: custom
    plugin_call: {plugin: relic_backend, function: nll, bindings: {p: DD_pvalue}}
\end{yaml}

Likelihood composition is kept on the model side: the core supplies the
kernels and decides neither the datasets nor which terms are active. Two
variants of a model --- normal and inverted ordering, say --- share every
parameter and matrix definition and import different likelihood files, and a
single constraint can be added, dropped or replaced without touching the rest
of the specification.

\subsection{Plugin interface}
\label{sec:plugins}

Every \code{plugin\_call} is dispatched through one global registry in the C++
core, keyed by the pair (plugin name, function name) and populated before a
model that references it is compiled. The registry does not distinguish
compiled from Python functions: \code{plugin\_call: \{plugin, function,
bindings, options\}} is identical YAML in both cases. Two routes populate it.

\paragraph{Compiled plugins.} A C++ plugin is one \code{.cpp} file under
\code{src/plugins/}, paired with a CMake fragment under \code{cmake/plugins/}
that the build discovers automatically. At library-load time it registers its
functions:
\begin{snippet}
bsm::core::Value omega(const bsm::core::PluginInvocation& invocation) {
  // invocation.arguments and invocation.options hold the bound values
  return /* ... call into the external backend ... */;
}
struct PluginRegistrar {
  PluginRegistrar() {
    bsm::core::register_plugin_function("micromegas", "omega", omega);
  }
};
const PluginRegistrar kRegistrar{};  // runs once, at static-initialisation time
\end{snippet}
A \code{PluginInvocation} carries the resolved \code{bindings:} and
\code{options:} as string$\to$\code{Value} maps, so a plugin sees plain real,
boolean, complex or string values, never YAML or graph internals. Building a
compiled plugin against its backend is opt-in at CMake configure time (one
boolean \code{BSM\_SCANNER\_BUILD\_<NAME>} per plugin, off by default, plus
backend paths). A backend with process-global state, as micrOMEGAs has, can be
linked against only one model per build.

\paragraph{Python plugins.}
\code{bsm\_scanner.register\_plugin\_function(plugin, function, callback)}
registers a callable \code{callback(arguments: dict, options: dict) -> value}
in the same registry. A model loads the registering module through the
top-level \code{python\_plugins:} key, resolved like \code{imports:} (relative
to the declaring fragment and deduplicated), so no driver script is needed:
\begin{snippet}
# my_plugin.py
import bsm_scanner
def quadruple(args, options):
    return args["value"] * 4.0
bsm_scanner.register_plugin_function("my_backend", "quadruple", quadruple)
\end{snippet}
\begin{yaml}
python_plugins:
  - my_plugin.py          # or a dotted, importable module name
observables:
  - {name: y, plugin_call: {plugin: my_backend, function: quadruple, bindings: {value: x}}}
\end{yaml}
This route suits backends that are themselves Python, light estimators, and
prototypes; it does not require the model author to touch C++ or the build
system. Each Python call crosses the C++/Python boundary, so a Python plugin
adds its own cost to every point at which it is evaluated.

\paragraph{Example: micrOMEGAs.} The package ships a generic \code{micromegas}
plugin (\code{src/plugins/micromegas.cpp}) registering seven functions ---
\code{omega}, \code{sigma\_si}, \code{dd\_pvalue}, \code{candidate\_mass},
\code{candidate\_valid}, \code{candidate\_name} and \code{target\_match}. It uses
only the public micrOMEGAs C interface (\code{assignValW},
\code{sortOddParticles}, \code{darkOmega}/\code{darkOmegaN},
\code{nucleonAmplitudes}, \code{DD\_pval}), and caches the results per point so
that several plugin-backed observables share one micrOMEGAs evaluation. The
CalcHEP model it uses is selected at configure time
(\code{BSM\_SCANNER\_MICROMEGAS\_ROOT}, \code{\_MODEL\_ROOT},
\code{\_CALCHEP\_ROOT}). No parameter name is matched automatically: each
\code{bindings:} entry is the author's explicit map from a CalcHEP parameter to
a model node, reused across observables with a YAML anchor:
\begin{yaml}
observables:
  - {name: Omega, value_type: real,
     plugin_call: {plugin: micromegas, function: omega,
                    bindings: &micromegas_bindings {MH01: MH1, Y1r1: g1er}}}
  - {name: SIxsec, value_type: real,
     plugin_call: {plugin: micromegas, function: sigma_si, bindings: *micromegas_bindings}}
likelihoods:
  - {name: Omega_term, kind: gaussian, observable: Omega, mean: 0.120, sigma: 0.0012}
\end{yaml}
The command
\begin{snippet}
bsm-scanner generate-dm-config --map map.yaml --dm-target-name '~o1' \
  --CalcHEP <model_root> --micromegas-root <root> --output-dir <dir>
\end{snippet}
generates the \code{plugin\_call} boilerplate and a CMake cache script from a
binding map supplied by the user.

\subsection{Outputs}

Any named quantity --- parameter, derived scalar, observable or per-term
likelihood contribution --- is written to \code{points.csv} by listing it under
\code{outputs:},
\begin{yaml}
outputs:
  save: [Mpsi, mphi, dm21, br_mu_e_gamma, ObliqueS, ObliqueT]
\end{yaml}
Only requested outputs are materialised, which keeps the active subgraph, and
therefore the cost per point, as small as the analysis requires.

%%%%%%%%%%%%%%%%%%%%%%%%%%%%%%%%%%%%%%%%%%%%%%%%%%%%%%%%%%%%%%%%%%%%%%%%%%%%%%%
\section{Scanning engines}
\label{sec:engines}

Section~\ref{sec:model} defines a single compiled objective.
Table~\ref{tab:engines} lists the four engines that can drive it and the two
optional post-scan stages. An engine is selected by one line,
\begin{yaml}
scan:
  engine: adaptive_diver
\end{yaml}
and every engine consumes the same compiled model, parameter ordering and
validity semantics, so that changing the engine changes nothing about the
physics being explored.

\begin{table}[t]
\centering\small
\begin{tabular}{lp{9.7cm}}
\toprule
Engine & Role \\
\midrule
\code{serial\_random} & Independent draws from the declared priors. No optimisation; the baseline, and the appropriate choice when the likelihood is a pure veto (section~\ref{sec:res_zprime}). \\
\code{de\_scipy} & \textsc{SciPy} differential evolution~\cite{Storn:1997de,Virtanen:2019joe}, an independent reference implementation behind the same interface. \\
\code{adaptive\_diver} & Native JADE-type~\cite{Zhang:2009jade} differential evolution with self-adapting $F$ and $CR$, optional local refinement of elite points, and convergence diagnostics. \\
\code{basin\_scan} & Orchestration of space-filling exploration, clustering of promising points, construction of focused boxes, and independent \code{adaptive\_diver} runs inside each box. \\
\midrule
posterior (\code{emcee}) & Optional local Markov-chain sampling around the best-fit or elite points of a completed scan~\cite{Foreman-Mackey:2013}. \\
statistics & Optional likelihood-weighted summaries and intervals from a completed scan's saved points. \\
\bottomrule
\end{tabular}
\caption{Scanning engines and post-scan stages.}
\label{tab:engines}
\end{table}

\subsection{Common settings}

All engines share a small set of fields:
\begin{yaml}
scan:
  engine: adaptive_diver
  seed: 12345
  save_every: 250
  settings:
    objective: nll             # or posterior_nll, adding -ln(prior)
    max_evaluations: 30000     # or engine-specific generation settings
    invalid_penalty: 1.0e12
    save_invalid_points: true
    verbose: 1
\end{yaml}
A point that fails a theory check or yields a non-finite objective is passed to
the engine with the value \code{invalid\_penalty}, so that the optimiser never
receives a NaN; the failure reason remains in \code{points.csv} and
\code{summary.json}. Because invalid points form a plateau at this value, a
population that contains no valid member has no gradient information; this is
the situation \code{basin\_scan} is designed for.
\code{serial\_random} requires \code{max\_evaluations};
\code{adaptive\_diver} requires either that or a generation count;
\code{de\_scipy} currently runs with a single worker.

\subsection{\texttt{adaptive\_diver}}

\code{adaptive\_diver} is the default for graded likelihoods. It uses the
current-to-$p$best mutation of JADE~\cite{Zhang:2009jade},
\begin{equation}
\bm{v}_i = \bm{x}_i + F_i(\bm{x}_{p{\rm best}}-\bm{x}_i) + F_i(\bm{x}_{r_1}-\bm{x}_{r_2}),
\end{equation}
where $\bm{x}_{p{\rm best}}$ is drawn from the best fraction $p$ of the
population, followed by binomial crossover with rate $CR_i$. The per-individual
$F_i$ and $CR_i$ are drawn around running means that are updated after every
generation from the values that produced successful trial vectors, and clipped
to the configured ranges:
\begin{yaml}
scan:
  engine: adaptive_diver
  adaptive_diver:
    population_size: 96
    max_generations: 300
    p_best_fraction: 0.2
    mutation: {adaptive: true, F_min: 0.05, F_max: 0.7, initial_mean: 0.35, learning_rate: 0.1}
    crossover: {adaptive: true, CR_min: 0.1, CR_max: 0.9, initial_mean: 0.55, learning_rate: 0.1}
    bounds: {handling: reflect}
    convergence: {patience: 60, population_std_tol: 1.0e-8, min_delta_chi2: 1.0e-8}
    local_refinement: {enabled: true, method: Powell, n_elites: 6, maxiter: 1200}
\end{yaml}
Table~\ref{tab:adaptive} summarises the options. The run terminates when the
generation or evaluation budget is exhausted (\code{max\_generations}), when the
best objective has not improved by more than \code{min\_delta\_chi2} for
\code{patience} generations (\code{patience}), or when the population spread
falls below \code{population\_std\_tol} (population collapse); the stop reason
is written to \code{summary.json}. Local refinement is applied to the elite
points after the global search has terminated; running it inside the
generational loop competes with the global search for budget.

\begin{table}[t]
\centering\small
\begin{tabular}{ll}
\toprule
Block & Controls \\
\midrule
\code{population\_size}, \code{max\_generations} & population and budget \\
\code{p\_best\_fraction} & fraction of the population eligible as $\bm{x}_{p{\rm best}}$ \\
\code{mutation}, \code{crossover} & ranges and learning rates of the adaptive $F$ and $CR$ means \\
\code{bounds.handling} & treatment of trial vectors outside the box (\code{reflect}, \dots) \\
\code{convergence} & patience and tolerances for early termination \\
\code{local\_refinement} & optional \textsc{SciPy} Powell or L-BFGS-B polish of elite points \\
\bottomrule
\end{tabular}
\caption{Options of \code{scan.adaptive\_diver} (alias \code{adaptive\_de}); the
shipped documentation gives the full schema.}
\label{tab:adaptive}
\end{table}

\subsection{\texttt{basin\_scan}}

\code{basin\_scan} is intended for problems in which the viable region occupies
a small, possibly disconnected fraction of the prior volume, as is common in
radiative neutrino-mass and dark-matter models with many simultaneous validity
conditions. It is an orchestration layer that sees only parameter names,
bounds, the objective and the validity flags. Its default workflow samples the
full box (Latin hypercube~\cite{McKay:1979lhs}, Sobol or uniform), selects
promising valid points by total objective or by a per-term balanced rule,
clusters them in normalised coordinates (DBSCAN~\cite{Ester:1996dbscan}),
builds a focused box from each cluster's quantiles, and runs
\code{adaptive\_diver} independently in each box:
\begin{yaml}
scan:
  engine: basin_scan
  basin_scan:
    exploration: {method: latin_hypercube, n_points: 50000}
    selection: {mode: top_fraction, top_fraction: 0.02, max_points: 2000}
    clustering: {method: dbscan, eps_fraction: 0.08, min_samples: 10, max_clusters: 8}
    boxes: {construction: quantile, padding_fraction: 0.25, min_width_fraction: 0.02}
    focused_engine: {name: adaptive_diver, population_size: 60, max_generations: 1500}
\end{yaml}
For very small viable fractions, the opt-in \code{progressive\_exploration} mode
repeats the cycle over several rounds, each sampled inside boxes built from the
previous round's selection and combined with a preserved elite archive. Further
opt-in stages compose with this workflow: guided sampling (below); staged
evaluation, which evaluates a reduced graph before an expensive plugin-backed
observable so that points failing an analytic cut do not pay for it; and a
refocusing stage that can shrink a focused box before the final
\code{adaptive\_diver} run.

\subsection{Guided sampling}

\code{guided\_sampling} transforms a sampled point before evaluation and is
useful when uniform sampling almost never reaches a narrow target region:
\begin{yaml}
guided_sampling:
  enabled: true
  stages:
    - name: balanced_yukawas
      type: complex_vector_norm
      probability: 0.5
      vectors:
        - components: {real: [yr1, yr2, yr3], imag: [yi1, yi2, yi3]}
          norm_range: [1.0e-5, 1.0]
          scale: log
\end{yaml}
The built-in stage types are Gaussian or discrete resampling of named
parameters (\code{prior\_profile}), sampling of a complex vector at a chosen
norm and random direction (\code{complex\_vector\_norm}), rescaling of a
parameter group (\code{parameter\_rescale}) and an arbitrary user Python
function (\code{point\_function}). The framework clips the result to the
declared bounds before evaluation. Guided sampling changes the proposal
distribution; samples obtained with it are not distributed according to the
declared priors and must not be used as such in a Bayesian interpretation.

\subsection{Posterior refinement and statistics}

An optimisation scan locates the best fit; it does not by itself provide
calibrated intervals. Two optional stages address this. Posterior refinement
samples the posterior locally around the best-fit or elite region with
\code{emcee}~\cite{Foreman-Mackey:2013}:
\begin{yaml}
posterior:
  enabled: true
  method: emcee
  start_from: {source: auto, n_walkers: 64, initialization: elite_covariance}
  mcmc: {n_steps: 20000, burn_in: 5000, thin: 10, seed: 12345}
\end{yaml}
It is a local sampler, intended to run after \code{basin\_scan} or another
global engine has located the relevant basin. The lighter \code{statistics:}
block computes likelihood-weighted summaries and intervals from the saved
valid points of an optimisation scan,
\begin{yaml}
statistics: {enabled: true, method: de_weighted, credible_levels: [0.68, 0.95]}
\end{yaml}
with weights $w_i\propto\exp[-\tfrac12(\chi^2_i-\chi^2_{\min})]=\exp[-(\Fobj_i-\Fobj_{\min})]$.
The points of an optimiser are not samples of the likelihood, so
these weights give a profile-like summary of the explored region, not a
posterior; calibrated intervals require the \code{emcee} stage.

%%%%%%%%%%%%%%%%%%%%%%%%%%%%%%%%%%%%%%%%%%%%%%%%%%%%%%%%%%%%%%%%%%%%%%%%%%%%%%%
\section{Examples}
\label{sec:examples}

We apply the full pipeline --- declaration, scan and best fit --- to six
benchmark models with $4$ to $26$ free parameters as listed in Table~\ref{tab:benchmarks}.
They are chosen to span the likelihood structures met in practice: a graded fit
to tabulated profiles, a degenerate plateau, a validity-dominated fit, an
exactly quadratic fit with flat directions, a pure veto problem, and a fit to
flavour anomalies with a degenerate optimum. The observables are tree-level or
leading-order analytic expressions documented with each model; they illustrate
the specification and scanning pipeline and are not intended as new
phenomenological results. For each model, the complete specification and the
scan configuration are distributed with the package. Table~\ref{tab:validation} collects the
analytic expectations against which the scan results are checked.

\begin{table}[t]
\centering\small
\begin{tabular}{clcl}
\toprule
\# & Model & Par. & Likelihood terms \\
\midrule
1 & Scotogenic~\cite{Ma:2006km}                     & 26 & 6 NuFIT tables, 1 Gaussian, 8 hard cuts \\
2 & Minimal $B\!-\!L$~\cite{Basso:2008iv}           & 7  & 1 Gaussian, 3 hard cuts \\
3 & Two-Higgs-doublet, Type~II~\cite{Branco:2011iw} & 6  & 2 Gaussian, 1 hard cut \\
4 & SMEFT, Warsaw basis~\cite{Grzadkowski:2010es}   & 9  & 1 bivariate Gaussian, 1 Gaussian \\
5 & Simplified $Z'$ dark matter~\cite{Abercrombie:2015wmb} & 4 & 2 hard cuts \\
6 & Scalar leptoquark $S_1$~\cite{Buchmuller:1986zs,Dorsner:2016wpm} & 4 & 1 bivariate Gaussian, 1 asym.\ Gaussian, 3 hard cuts \\
\bottomrule
\end{tabular}
\caption{Benchmark models. Parameter counts and likelihood terms are those of
the specifications scanned in this section; theory checks are not counted.}
\label{tab:benchmarks}
\end{table}

\subsection{Scotogenic model}
\label{sec:res_scotogenic}

The Scotogenic model~\cite{Ma:2006km} generates neutrino masses at one loop
through an inert scalar doublet $\eta$ and three $Z_2$-odd singlet fermions
$N_k$,
\begin{equation}
(M_\nu)_{\alpha\beta}=\sum_{k} h_{\alpha k}h_{\beta k}\,\Lambda_k ,
\end{equation}
with $\Lambda_k$ the loop function of section~\ref{sec:parameters}. The
specification scans $26$ real parameters: the inert-doublet mass and quartic
couplings, the three singlet masses, and the complex $3\times3$ Yukawa matrix
$h$ in real and imaginary parts. The likelihood consists of the six NuFIT
normal-ordering profiles ($\ssq{12}$, $\ssq{13}$, $\ssq{23}$, $\dCP$,
$\dmsq{21}$, $\dmsq{3\ell}$), a Gaussian on the oblique parameter $T$, and hard
cuts on $\sum m_\nu<0.12\eV$, on $\mu\to e\gamma$~\cite{MEGII:2025gzr},
$\tau\to e\gamma$~\cite{BaBar:2009hkt} and $\tau\to\mu\gamma$~\cite{Belle:2021ysv},
and on the LEP limits on the $Z$ and $W$ widths and on the charged and neutral
inert scalars~\cite{Pierce:2007ut}. The neutrino mass matrix is diagonalised
with \code{svd\_complex}, which determines the masses, mixing angles and $\dCP$.

With population $96$ and at
most $300$ generations ($\simeq3.3\times10^{4}$ evaluations) \code{adaptive\_diver} reached a best objective of 0.64. It is important to mention that this 26-dimensional likelihood surface has many separated local minima, therefore a single run is not evidence of a global minimum; \code{basin\_scan} with progressive exploration is the engine designed for this situation.

\begin{figure}[t]
    \centering
    \subfloat[][]{\includegraphics[width=0.5\linewidth]{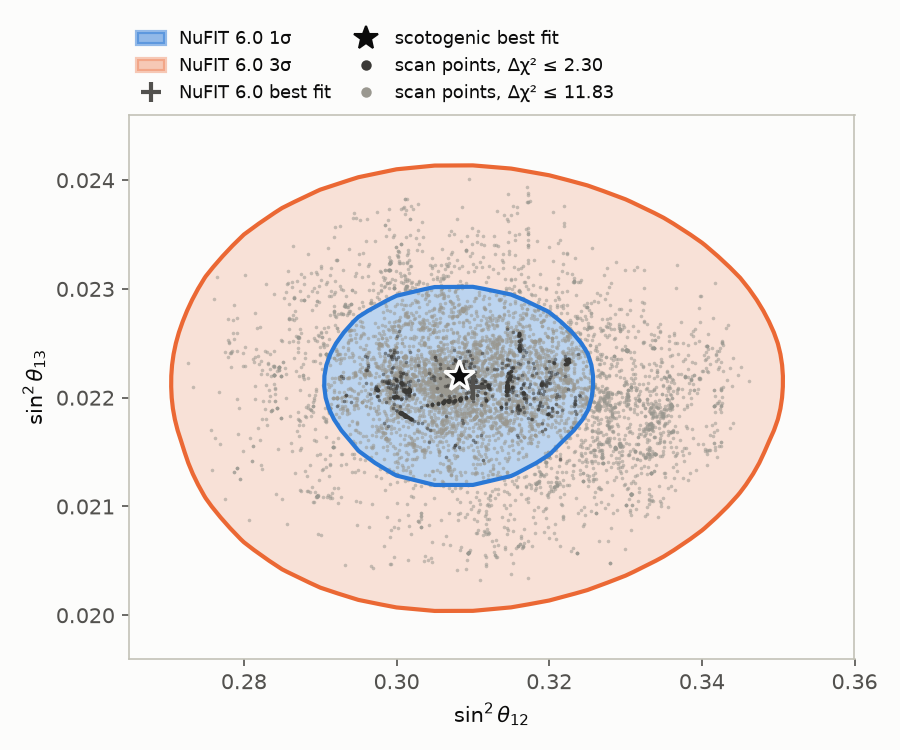}}
    \subfloat[][]{\includegraphics[width=0.5\linewidth]{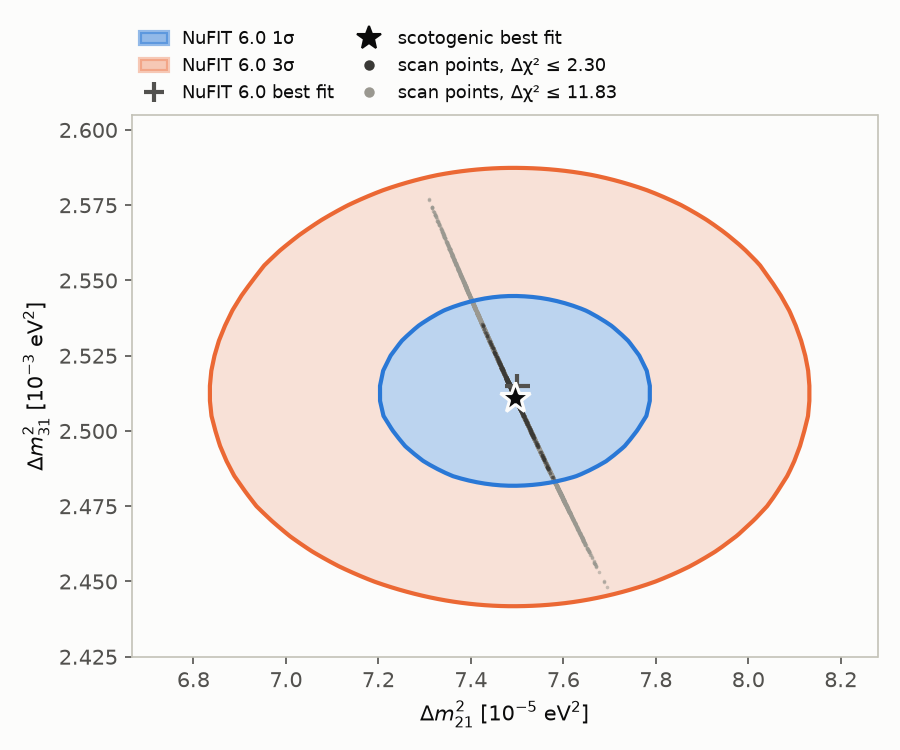}}
    \caption{Scotogenic model, normal ordering. \emph{(a)}
    $(\ssq{12},\ssq{13})$ plane; \emph{(b)} $(\dmsq{21},\dmsq{31})$ plane. The
    NuFIT-6.0 $1\sigma$ and $3\sigma$ regions are shown with the scan
    points and the best fit.}
    \label{fig:scotogenic}
\end{figure}

At the best point the objective is dominated by the atmospheric angle
($f_{\ssq{23}}=0.56$), followed by $\dCP$ ($0.11$) and $T$ ($0.013$); the
$\ssq{12}$, $\ssq{13}$, $\dmsq{21}$ and $\dmsq{3\ell}$ terms are at the minima of
their profiles.
Figure~\ref{fig:scotogenic} shows the scan points and the best fit in the
$(\ssq{12},\ssq{13})$ and $(\dmsq{21},\dmsq{31})$ planes.

\subsection{Minimal \texorpdfstring{$B\!-\!L$}{B-L} model}
\label{sec:res_bl}

The minimal $B\!-\!L$ model~\cite{Basso:2008iv} gauges $U(1)_{B-L}$, with a
$Z'$, a symmetry-breaking scalar and three right-handed neutrinos, and is
scanned over seven parameters: $\gBL$, $\vBL$, the heavy-scalar mass $m_{h_2}$,
the scalar mixing $\sin\alpha$ and three Majorana Yukawas $y_{N_i}$. Three
likelihood terms are hard cuts. The LEP-II contact-interaction
bound~\cite{Carena:2004xs} $\MZp/\gBL\ge7\TeV$ reduces, with
$\MZp=2\gBL\vBL$, to $\vBL\ge3.5\TeV$ independently of $\gBL$. The narrow-width
requirement $\Gamma_{Z'}/\MZp<0.1$ uses the full tree-level width summed over
quarks, charged leptons and light and heavy Majorana neutrinos with mass
thresholds, and restricts $\gBL\lesssim0.69$--$0.76$ depending on the number of
open heavy-neutrino channels. The third cut applies the ATLAS $139\,{\rm fb}^{-1}$
dilepton resonance search~\cite{ATLAS:2019erb} through the ratio of the
predicted $\sigma\times\mathrm{BR}(Z'\to ee)$ to the limit. The fourth term is
a Gaussian on the Higgs signal strength, $\mu=1.025\pm0.06$, the average of the
ATLAS and CMS Run-2 combinations~\cite{ATLAS:2022vkf,CMS:2022dwd}, with
$\mu=\cos^2\alpha\,(1-\mathrm{BR}_{\rm exo})$ and $\mathrm{BR}_{\rm exo}$ the
branching ratio of $h_1\to NN$ and $h_1\to Z'Z'$.

\begin{figure}[t]
\centering
\includegraphics[width=0.62\textwidth]{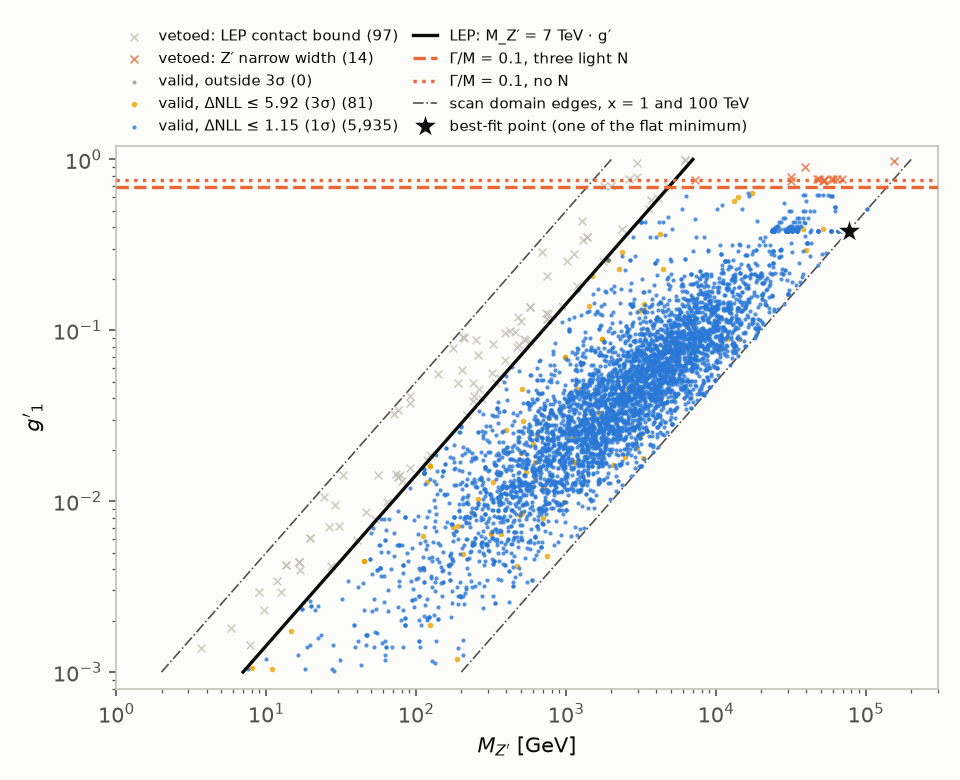}
\caption{Minimal $B\!-\!L$ model, $(\MZp,\gBL)$ plane. The diagonal line is the
LEP contact-interaction bound $\MZp=7\TeV\cdot\gBL$; the horizontal lines mark
$\Gamma_{Z'}/\MZp=0.1$ with three light and with no heavy neutrinos; the
dash-dotted lines are the edges of the scanned domain $\vBL\in[1,100]\TeV$.
Crosses are points vetoed by the LEP bound (grey) or the width cut (orange).}
\label{fig:bl}
\end{figure}

Since $\mu\le1$ in the model, the Gaussian is minimised at $\sin\alpha=0$ and
$\mathrm{BR}_{\rm exo}=0$, giving the analytic floor
$\Fobj_{\min}=\tfrac12(0.025/0.06)^2=0.087$ for any allowed
$(\gBL,\vBL,m_{h_2},y_N)$. The optimum is therefore a six-dimensional plateau,
and the scan serves to map the allowed region rather than to locate a point.
\code{adaptive\_diver} (population 128) reaches $\Fobj=0.087$ and terminates by
population collapse after $6.1\times10^3$ evaluations ($98\%$ valid). Figure~\ref{fig:bl} shows the points
fill the allowed wedge of the $(\MZp,\gBL)$ plane and
saturate the LEP bound to $0.15\%$ ($\min\MZp/\gBL=7.01\TeV$), confirming that
the cut is applied at its stated boundary. The Higgs term alone gives
$|\sin\alpha|<0.20$ at $1\sigma$ , while $m_{h_2}$ and the
$y_{N_i}$ are unconstrained over their ranges. 

\subsection{Two-Higgs-doublet model}
\label{sec:res_2hdm}

The CP-conserving Type~II two-Higgs-doublet model with softly broken
$Z_2$~\cite{Branco:2011iw} is scanned over six physical-basis parameters:
$m_H$, $m_A$, $m_{H^\pm}$, $\tan\beta$, $\cos(\beta-\alpha)$ and the soft-breaking
scale, parametrised as $\delta\equiv(m_H^2-M^2)\tan^2\beta/v^2$ with
$M^2=m_{12}^2/(s_\beta c_\beta)$. The quartic couplings follow from the
physical-basis inversion and are subject to the exact boundedness-from-below
conditions~\cite{Branco:2011iw} and to tree-level perturbative unitarity,
$|e_i|<8\pi$ for the eigenvalues $e_i$ of the $2\to2$ scattering matrix. The likelihood combines a Gaussian on the
oblique parameter $T$ computed exactly at one loop, $T=0.00\pm0.06$, the Type~II bound
$m_{H^\pm}>800\GeV$ from $\bar B\to X_s\gamma$~\cite{Misiak:2020vlo} as a hard
cut, and a Gaussian on $\kappa_V=\sin(\beta-\alpha)$,
$\kappa_V=1.035\pm0.031$~\cite{ATLAS:2022vkf}.

\begin{figure}[t]
\centering
\subfloat[][]{\includegraphics[width=0.5\textwidth]{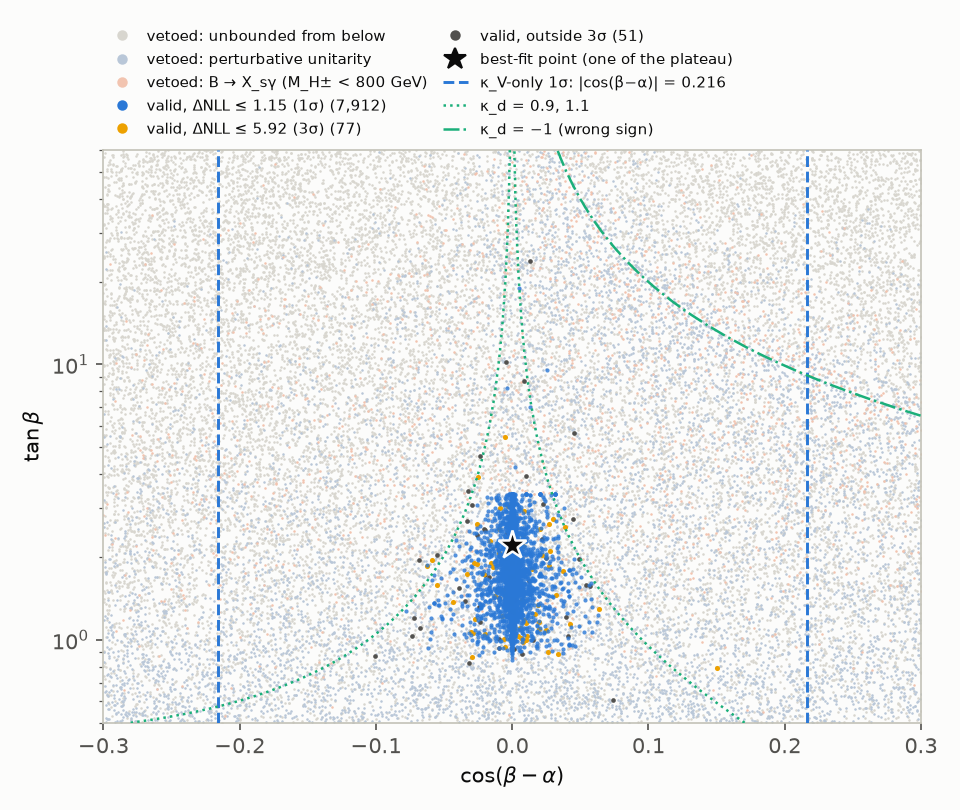}}
\subfloat[][]{\includegraphics[width=0.5\textwidth]{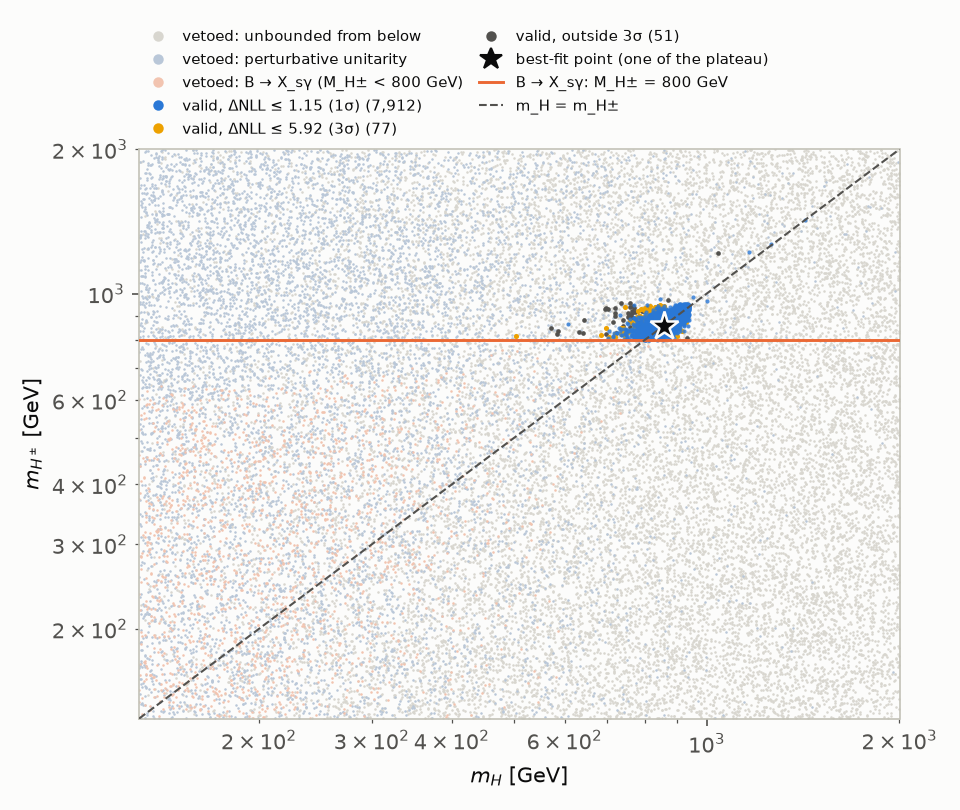}}
\caption{Type~II two-Higgs-doublet model. \emph{(a)}
$(\cos(\beta-\alpha),\tan\beta)$ plane. The dashed lines mark the $1\sigma$ edge from the
$\kappa_V$ term alone, and the dotted and dash-dotted curves are the Type~II
contours $\kappa_d=0.9,\,1.1$ and the wrong-sign branch $\kappa_d=-1$.
\emph{(b)} $(m_H,m_{H^\pm})$ plane: valid points lie near the custodial diagonal
and above the $\bar B\to X_s\gamma$ bound $m_{H^\pm}=800\GeV$; background points
are vetoed by stability, unitarity or $\bar B\to X_s\gamma$.}
\label{fig:2hdm}
\end{figure}

This benchmark is dominated by its validity boundary. Stability and unitarity
confine $M^2$ to a window of width ${\sim}v^2/\tan^2\beta$ below $m_H^2$, so
that uniform sampling in $m_{12}^2$ yields a valid fraction of
$5\times10^{-6}$, and $1.4\times10^{-4}$ in $\delta$. A randomly initialised
\code{adaptive\_diver} population then contains no valid member, sees the flat
invalid plateau, and stalls. \code{basin\_scan} resolves this: a
$4\times10^5$-point Latin-hypercube exploration supplies 50 valid seeds, and the
seeded \code{adaptive\_diver} stage reaches a valid fraction of $93\%$ and
terminates by population collapse after $8.6\times10^3$ evaluations at
$\Fobj=0.637$, the analytic floor $\tfrac12(0.035/0.031)^2$ set by
$\kappa_V\le1$. Among near-optimal points ($\Delta\Fobj\le10^{-4}$), every point
has $m_{H^\pm}$ within $20\GeV$ of $m_A$ or $m_H$, with median splittings
$|m_{H^\pm}-m_A|=1.1\GeV$ and $|m_{H^\pm}-m_H|=1.9\GeV$, and the median
$|\cos(\beta-\alpha)|$ is $5\times10^{-5}$ as shown in Figure~\ref{fig:2hdm}. These are
the custodial and alignment limits of the model, neither of which is imposed in
the specification: the $T$ term enforces the mass degeneracy and the
$\kappa_V$ term enforces alignment. Within the $1\sigma$ region the alignment
cone widens to $|\cos(\beta-\alpha)|\lesssim0.08$ for $\tan\beta\lesssim3.4$; it
is bounded by stability and unitarity rather than by the Higgs data alone,
which allow $|\cos(\beta-\alpha)|<0.15$, because the theory constraints tighten
as $\cos(\beta-\alpha)\tan\beta$ grows.

\subsection{SMEFT in the Warsaw basis}
\label{sec:res_smeft}

The Warsaw-basis SMEFT benchmark~\cite{Grzadkowski:2010es} scans nine Wilson
coefficients ($C_{H\Box}$, $C_{HD}$, $C_{HWB}$, $C_{HG}$, $C_{HW}$, $C_{HB}$ and
the third-generation $C_{uH}$, $C_{dH}$, $C_{eH}$) at fixed scale $\Lambda=1\TeV$. The
cutoff is not scanned, since every observable depends only on $C_i/\Lambda^2$
and a free $\Lambda$ would be an exact flat direction. The observables are
tree-level, linear maps: the oblique parameters~\cite{Han:2004az}
\begin{equation}
S=\frac{4s_Wc_W}{\alpha}\,\epsilon\,C_{HWB},\qquad
T=-\frac{\epsilon}{2\alpha}\,C_{HD},\qquad \epsilon=\frac{v^2}{\Lambda^2},
\end{equation}
and the Higgs coupling modifiers, including the $hgg$ and $h\gamma\gamma$
contact terms~\cite{Brivio:2017vri}. The likelihood combines the correlated
PDG~2024 $S$--$T$ fit (section~\ref{sec:likelihoods}) with
$\kappa_V=1.035\pm0.031$~\cite{ATLAS:2022vkf}; two theory checks impose
$\epsilon<0.1$ and the linear-EFT condition $|\delta\kappa|<0.5$ on every
coupling modifier, which $39\%$ of the prior volume satisfies.
\code{adaptive\_diver} terminates by population collapse after
$1.5\times10^4$ evaluations ($95\%$ valid) at $\Fobj=2\times10^{-12}$. The
minimum is zero because three measurements fix three independent combinations:
$C_{HWB}=-3.06\times10^{-3}$ and $C_{HD}=-2.58\times10^{-3}$ reproduce the central
$S$ and $T$, in agreement with the analytic inversion to four digits, and
$C_{H\Box}-C_{HD}/4=0.58$ reproduces $\kappa_V$. The remaining six coefficients
span a degenerate manifold bounded only by the validity condition.

\begin{figure}[t]
\centering
\subfloat[][]{\includegraphics[width=0.5\textwidth]{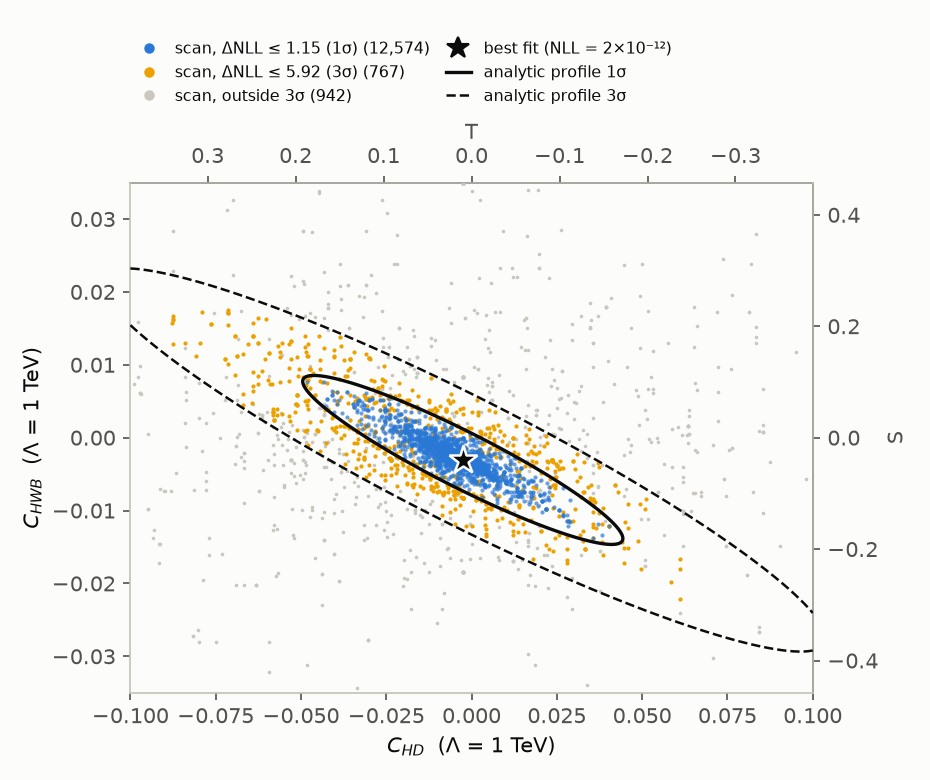}}
\subfloat[][]{\includegraphics[width=0.45\textwidth]{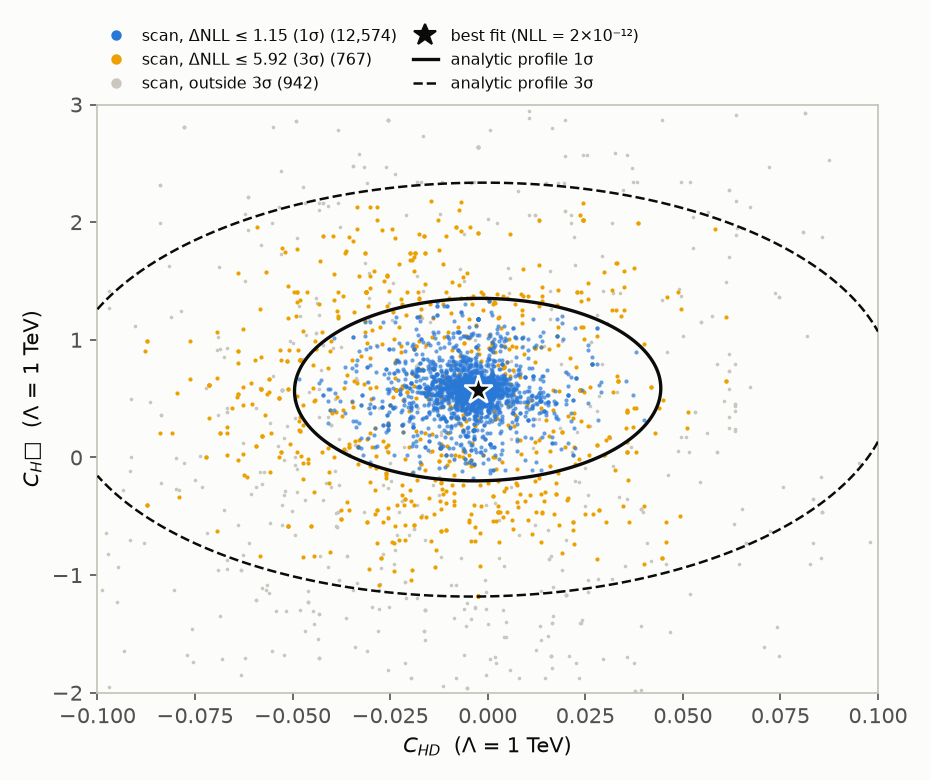}}
\caption{SMEFT benchmark at $\Lambda=1\TeV$: valid scan points coloured by
$\Delta\Fobj$ ($1\sigma$ and $3\sigma$ for two parameters), analytic
profile-likelihood contours ($1\sigma$ solid, $3\sigma$ dashed) and the best fit
($\star$). \emph{(a)} $(C_{HD},C_{HWB})$ plane, with the corresponding $T$ (top)
and $S$ (right) axes: the PDG $S$--$T$ ellipse mapped onto the Wilson
coefficients. \emph{(b)} $(C_{HD},C_{H\Box})$ plane, in which $C_{H\Box}$ is
fixed by $\kappa_V$ and $C_{HD}$ by $T$.}
\label{fig:smeft}
\end{figure}

Figure~\ref{fig:smeft} show the $(C_{HD},C_{HWB})$ plane. Its strong correlation constrains
$C_{HWB}+0.25\,C_{HD}$ to $\pm2.9\times10^{-3}$, while the one-dimensional
$1\sigma$ ranges are $C_{HWB}\in[-0.010,0.004]$ and $C_{HD}\in[-0.031,0.026]$.
The universal coefficient $C_{H\Box}\in[0.07,1.06]$ is diluted by
$\epsilon\simeq0.06$.

\subsection{Simplified \texorpdfstring{$Z'$}{Z'} dark matter}
\label{sec:res_zprime}

The simplified $Z'$ model~\cite{Abercrombie:2015wmb} is the leptophobic
vector-mediator benchmark of the Dark Matter Forum, with Dirac dark matter and a
universal quark coupling, scanned over $\MZp$, $m_\chi$, $g_q$ and $g_\chi$. The
mediator width follows the Forum's minimal-width expression with all quark
thresholds; neglecting the top-quark threshold would overestimate it by $17\%$
below $\MZp=2m_t$. Two hard cuts form the likelihood: the narrow-width
condition $\Gamma_{Z'}/\MZp<0.3$, and the LZ $90\%$ CL spin-independent
limit~\cite{LZ:2024zvo} applied to
\begin{equation}
\sigma_{\rm SI}=\frac{9g_q^2g_\chi^2\mu_{n\chi}^2}{\pi\MZp^4}
\end{equation}
\cite{Boveia:2016mrp}.

\begin{figure}[t]
\centering
\includegraphics[width=0.62\textwidth]{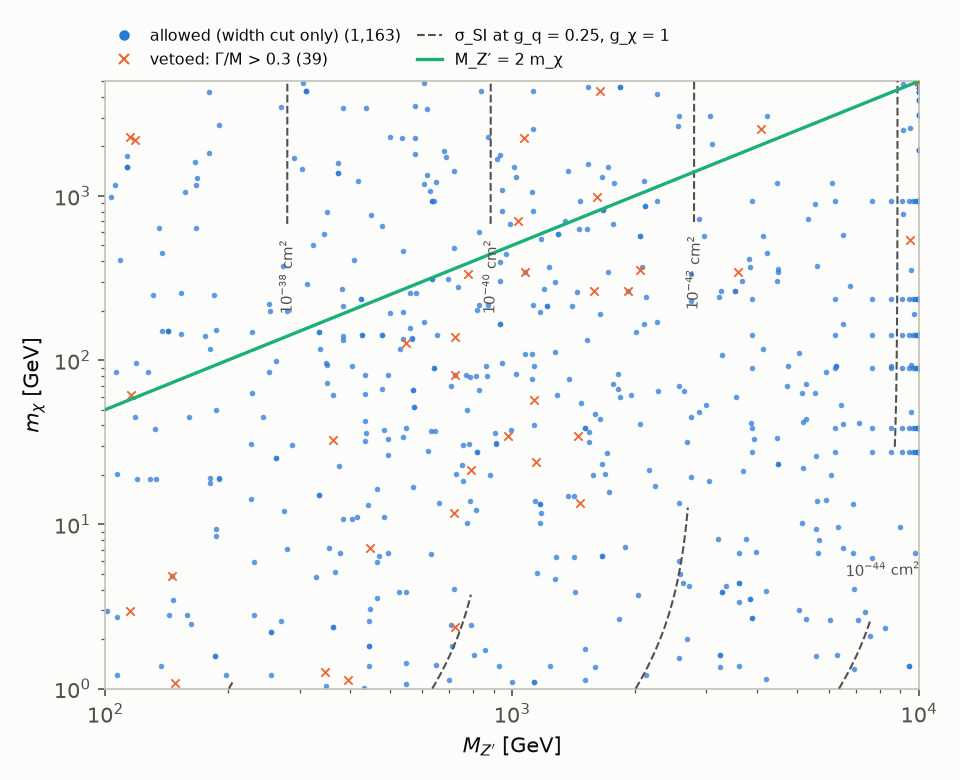}
\caption{Simplified $Z'$ model, $(\MZp,m_\chi)$ plane. Points allowed by the
width condition (circles) cover the full prior range; crosses are vetoed by
$\Gamma_{Z'}/\MZp>0.3$, which is driven by the couplings. Dashed lines are
contours of $\sigma_{\rm SI}$ at the Forum benchmark couplings ($g_q=0.25$,
$g_\chi=1$), from $10^{-38}$ to $10^{-44}\,$cm$^2$; the solid line marks the
invisible-decay threshold $\MZp=2m_\chi$.}
\label{fig:zprime}
\end{figure}

Every valid point has $\Fobj=0$, so there is no optimum to locate: this is the
case, noted in section~\ref{sec:engines}, for which the figure of merit is the
coverage of the allowed region. \code{adaptive\_diver} shows the expected
signature: its initial population already sits at $\Fobj=0$ and it terminates by
population collapse after three generations ($1.2\times10^3$ evaluations,
$97\%$ valid); \code{serial\_random} is the appropriate engine here. The width
cut constrains the couplings rather than the masses: for $\MZp\gg2m_t,2m_\chi$,
$\Gamma_{Z'}/\MZp=(18g_q^2+g_\chi^2)/(12\pi)$, so points with
$18g_q^2+g_\chi^2\gtrsim3.6\pi$ (i.e.\ $g_q\gtrsim0.8$) are vetoed and the full
$(\MZp,m_\chi)$ range survives is shown in Figure~\ref{fig:zprime}. The mass dependence
enters through direct detection: at fixed couplings $\sigma_{\rm SI}$ is
independent of $m_\chi$ for $m_\chi\gg m_n$ and scales as $\MZp^{-4}$, so a
Xenon limit removes a slice of light mediators whose position moves with
$g_qg_\chi$.

\subsection{Scalar leptoquark \texorpdfstring{$S_1$}{S1}}
\label{sec:res_lq}

The weak-singlet scalar leptoquark $S_1\sim(\bar3,1,1/3)$ of the
Buchm\"uller--R\"uckl--Wyler classification~\cite{Buchmuller:1986zs,Dorsner:2016wpm}
is scanned over its mass $M_{S_1}\in[0.5,10]\TeV$ and three real couplings
$y_L^{33}$, $y_L^{23}$ and $y_R^{23}$. At tree level these generate
$b\to c\tau\bar\nu$ through $g_{V_L}$ and $g_{S_L}=-4g_T$~\cite{Angelescu:2018tyl},
evolved to $m_b$~\cite{Blanke:2018yud}, and $b\to s\nu\bar\nu$ through a
left-handed $(\bar sb)(\bar\nu\nu)$ operator~\cite{Dorsner:2016wpm}. The
likelihood consists of the correlated HFLAV average of $R(D)$ and
$R(D^*)$~\cite{HFLAV:2022esi}, the Belle~II measurement
$\mathcal B(B^+\to K^+\nu\bar\nu)=(2.3^{+0.7}_{-0.6})\times10^{-5}$~\cite{Belle-II:2023esi}
as an asymmetric Gaussian, and three hard cuts: the $B_c$-lifetime
bound~\cite{Alonso:2016oyd}, a narrow-width condition $\Gamma_{S_1}/M_{S_1}<0.2$,
and the LHC pair-production limit.

\begin{figure}[t]
\centering
\subfloat[][]{\includegraphics[width=0.5\textwidth]{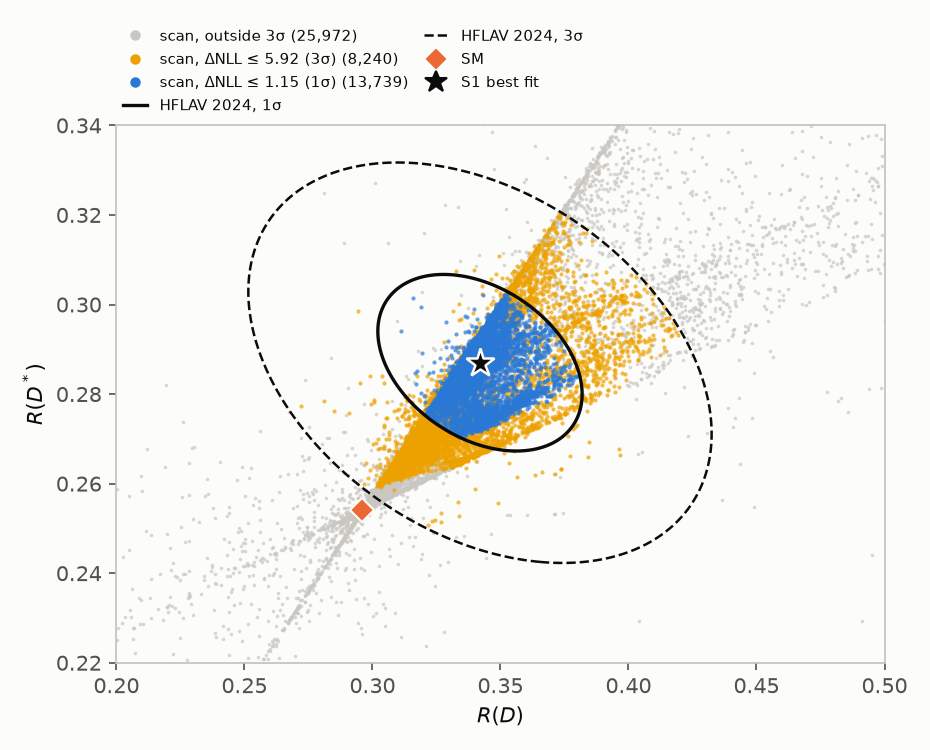}}
\subfloat[][]{\includegraphics[width=0.5\textwidth]{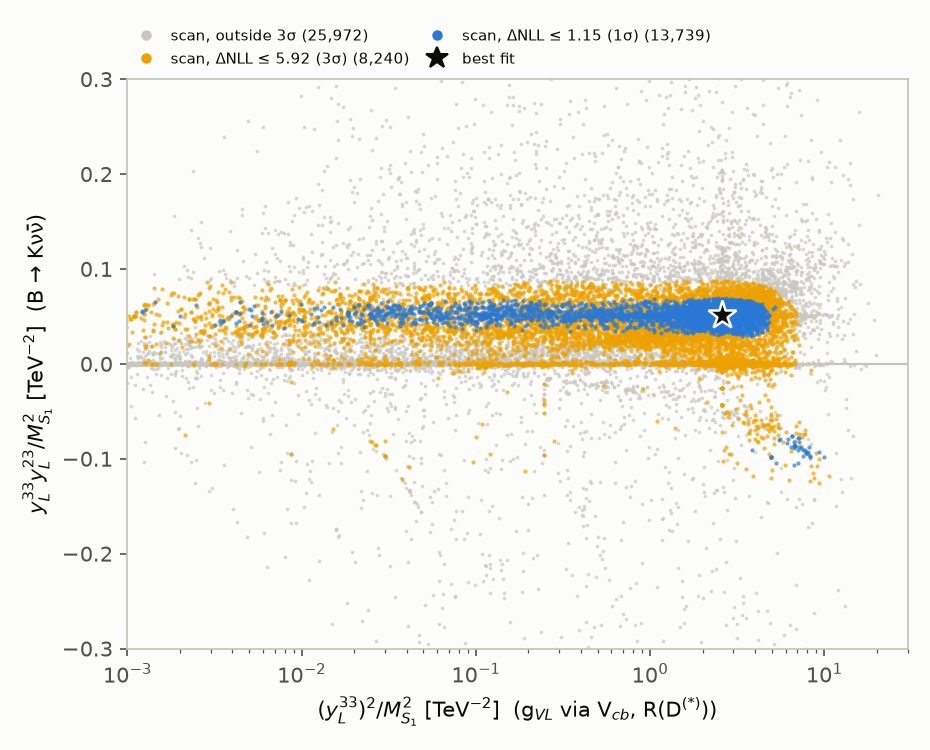}} \\
\subfloat[][]{\includegraphics[width=0.5\textwidth]{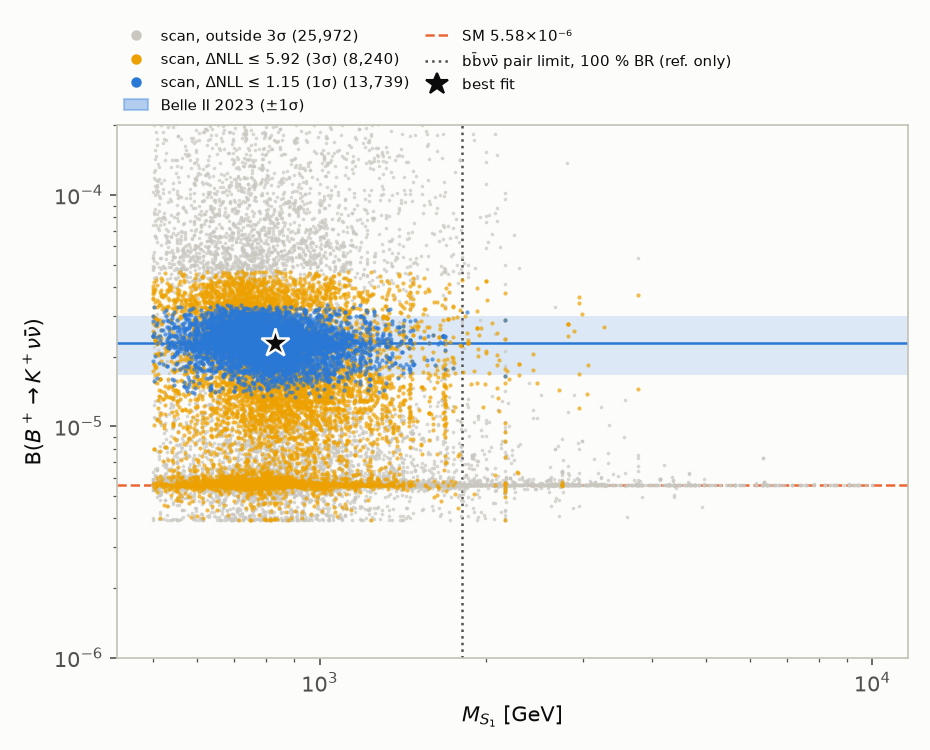}}
\caption{$S_1$ leptoquark without the LHC pair-production term: valid scan points
coloured by $\Delta\Fobj$ ($1\sigma$ and $3\sigma$ for two parameters).
\emph{(a)} $(R(D),R(D^*))$ plane with the HFLAV $1\sigma$ and $3\sigma$ contours,
the SM prediction and the best fit. \emph{(b)} The coupling combinations
$(y_L^{33})^2/M_{S_1}^2$, which drives $R(D^{(*)})$, and
$y_L^{33}y_L^{23}/M_{S_1}^2$, which drives $B\to K\nu\bar\nu$: the constructive
branch at $+0.05\TeV^{-2}$ and the overshooting branch at $-0.1\TeV^{-2}$.
\emph{(c)} $\mathcal B(B^+\to K^+\nu\bar\nu)$ against $M_{S_1}$, with the
Belle~II band, the SM value and the pair-production limit for
$\mathcal B(S_1\to b\nu)=100\%$, shown for reference.}
\label{fig:lq}
\end{figure}

The \code{adaptive\_diver} reaches
$\Fobj=1.7\times10^{-11}$ after $5.2\times10^4$ evaluations ($93\%$ valid),
reproducing both measurements at their central values. A light, strongly coupled
leptoquark with $M_{S_1}=0.83\TeV$ and $y_L^{33}=-1.34$ accommodates
$R(D^{(*)})$ mainly through the $V_{cb}(y_L^{33})^2$ term in $g_{V_L}$, while
$y_L^{33}y_L^{23}>0$ interferes constructively with the SM amplitude in
$b\to s\nu\bar\nu$ and enhances $\mathcal B(B^+\to K^+\nu\bar\nu)$ by a factor
$4.5$. The observables depend only on $(y_L^{33})^2/M_{S_1}^2$,
$y_L^{33}y_L^{23}/M_{S_1}^2$ and $y_L^{33}y_R^{23}/M_{S_1}^2$, so the optimum is a
line along which the couplings grow with the mass, terminated only by the width
condition: near-optimal points span $M_{S_1}=0.64$--$0.93\TeV$, the $1\sigma$
region extends to $1.7\TeV$ as shown in Figure~\ref{fig:lq}.
Essentially the whole preferred region lies below the
$b\bar b\nu\bar\nu$ pair-production limit of 1.8\,TeV, so the LHC term is expected to exclude most of it.

\subsection{Cross-checks}
\label{sec:validation}

Wherever the minimum or a constraint boundary is known analytically, the scan
result can be compared with it. Table~\ref{tab:validation} collects these
comparisons. They test the complete chain --- specification, compilation,
likelihood kernels and engine --- rather than isolated components; the
component-level tests (diagonalisation, mixing-matrix conventions, likelihood
kernels and the table edge treatment) are part of the package's test suite.

\begin{table}[t]
\centering\small
\begin{tabular}{llll}
\toprule
Model & Quantity & Analytic & Scan \\
\midrule
$B\!-\!L$ & $\Fobj_{\min}=\tfrac12(0.025/0.06)^2$ & $0.0868$ & $0.087$ \\
$B\!-\!L$ & $\min \MZp/\gBL$ (LEP cut boundary) & $7.00\TeV$ & $7.01\TeV$ \\
2HDM & $\Fobj_{\min}=\tfrac12(0.035/0.031)^2$ & $0.637$ & $0.637$ \\
2HDM & $|\cos(\beta-\alpha)|$, $1\sigma$, $\kappa_V$ term alone & $0.15$ & $0.15$ \\
SMEFT & $\Fobj_{\min}$ & $0$ & $2\times10^{-12}$ \\
SMEFT & $C_{HWB}$, $C_{HD}$ from central $S$, $T$ & $-3.06$, $-2.58\ (\times10^{-3})$ & agree to 4 digits \\
SMEFT & $1\sigma$ region in $(C_{HD},C_{HWB})$ & profile contour & coincident \\
$Z'$ & $g_q$ veto threshold at $g_\chi\to0$ & $0.79$ & $\simeq0.8$ \\
$S_1$ & $\Fobj_{\min}$ (2 measurements, 4 parameters) & $0$ & $1.7\times10^{-11}$ \\
\bottomrule
\end{tabular}
\caption{Comparison of scan results with analytic expectations.}
\label{tab:validation}
\end{table}

%%%%%%%%%%%%%%%%%%%%%%%%%%%%%%%%%%%%%%%%%%%%%%%%%%%%%%%%%%%%%%%%%%%%%%%%%%%%%%%
\section{Conclusions}
\label{sec:conclusions}

BSMScanner turns a BSM parameter scan from a program into a document. A model is
declared once in YAML --- parameters and priors, mass matrices, observables,
theory checks and likelihood terms --- and compiled into a typed C++ kernel
that every scanning engine drives through the same interface, at a cost of
$\mathcal{O}(10$--$50)\,\mu$s per point for the benchmarks considered here. The
model-independent machinery that private codes reimplement --- PMNS and CKM
construction, Takagi and singular-value diagonalisation, mixing-angle and phase
conventions, tabulated oscillation likelihoods --- is provided by the framework
and imported by name, and external calculations enter through a C++ or Python
plugin interface without breaking the declarative description.

The six benchmarks illustrate what this provides in practice. Changing the
physics --- replacing a proxy constraint by a measurement, adding a correlated
likelihood or reparametrising a model --- is an edit of a YAML entry, and
changing the search strategy is one line of the \code{scan:} section. This
makes it straightforward to match the engine to the structure of each problem:
\code{adaptive\_diver} for graded fits such as the SMEFT and leptoquark models;
\code{basin\_scan} for the two-Higgs-doublet model, whose valid region occupies
$10^{-4}$ of the prior volume and stalls a randomly initialised search; and
coverage rather than best objective for the plateau and pure-veto problems of
the $B\!-\!L$ and $Z'$ models.
Because every run records its seed, resolved configuration and framework
version, each result can be reproduced and audited from its specification.

%%%%%%%%%%%%%%%%%%%%%%%%%%%%%%%%%%%%%%%%%%%%%%%%%%%%%%%%%%%%%%%%%%%%%%%%%%%%%%%
\acknowledgments

%%%%%%%%%%%%%%%%%%%%%%%%%%%%%%%%%%%%%%%%%%%%%%%%%%%%%%%%%%%%%%%%%%%%%%%%%%%%%%%
\appendix

\section{Installation and command-line interface}
\label{app:availability}

BSMScanner is released under the MIT licence at
\url{https://github.com/mbelfkir/bsmscanner} and on the Python Package Index as
\code{bsm-scanner} (\url{https://pypi.org/project/bsm-scanner/}). It is
installed with \code{pip install bsm-scanner}; the compiled core is built
automatically. Table~\ref{tab:cli} lists the commands.

\begin{table}[h]
\centering\small
\begin{tabular}{lp{7cm}}
\toprule
Command & Action \\
\midrule
\code{bsm-scanner run ---model <file> ---run-dir <dir>} & run the scan declared in the model, writing the artefacts of Table~\ref{tab:artefacts} to \code{<dir>} \\
\code{bsm-scanner new-model <name>} & create a runnable starter model wired to the core library \\
\code{bsm-scanner core list} & list the reusable \code{core:} blocks \\
\code{bsm-scanner core show <block>} & print the constants, functions, observables and checks contributed by a block \\
\code{bsm-scanner generate-dm-config ...} & generate \code{plugin\_call} YAML and a CMake cache script for the micrOMEGAs plugin from a user-supplied binding map \\
\code{bsm-scanner ---version} & print the installed version \\
\bottomrule
\end{tabular}
\caption{Command-line interface.}
\label{tab:cli}
\end{table}

\bibliography{refs}

\end{document}